\documentclass[]{aa}
\usepackage{graphicx} % Required for inserting images
\usepackage{natbib}
\usepackage{xcolor}
\bibstyle{aa.bst}
\begin{document}
\title{
The curious case of 2MASS J15594729+4403595, an ultra-fast M2 dwarf with possible Rieger cycles
}
\author{S. Messina\inst{1}\fnmsep\thanks{Corresponding author:
  \email{sergio.messina@inaf.it}\newline}
\and G. Catanzaro\inst{1}
\and A. F. Lanza\inst{1}
\and D.~Gandolfi\inst{3}
\and  M.M. Serrano\inst{2}
    \and H.\,J.~Deeg\inst{5}
\and D. Garc\'ia-Alvarez\inst{6}
}
\offprints{Sergio Messina}
\institute{INAF-Catania Astrophysical Observatory, via S.Sofia, 78 I-95127 Catania, Italy \\
\email{sergio.messina@inaf.it}
\and   
Mizar Observatory, Madrid, Spain
\and Dipartimento di Fisica, Universit\'a degli Studi di Torino, via Pietro Giuria 1, I-10125, Torino, Italy\label{LSW} 
\and Departamento de Astrof\'\i sica, Universidad de La Laguna, 38206 La Laguna, Spain\label{LaLaguna}
\and Instituto de Astrofisica de Canarias, 38205, La Laguna, Tenerife, Spain 
}

\date{}
\titlerunning{Rieger cycles on 2MASS J15594729+4403595}
\authorrunning{S.\,Messina et al.}
\abstract {RACE-OC (Rotation and ACtivity Evolution in Open Clusters) is a project aimed at characterising the rotational and magnetic activity properties of the late-type members of open clusters, stellar associations, and moving groups of different ages. 
The evolution in time of rotation and activity at different masses sheds light on the evolution of the stellar internal structure, on magneto-hydrodynamic processes operating in the stellar interior, and on the coupling and decoupling mechanisms between the radiative core and the external convective envelope.
As part of this project, in the present paper we present the results of an investigation of a likely member of the AB Doradus association, the M-type star 2MASS J15594729+4403595.}
{In the present study, we aim to reveal the real nature of our target, which turned out to be a hierarchical triple system, to derive the stellar rotation period and surface differential rotation, and to characterise its photospheric magnetic activity.} {We have collected radial velocity and photometric time series, complemented with archive data, to determine the orbital parameters and the rotation period and we have used the spot modelling technique to  explore what causes its photometric variability. \rm } {We found 2MASS J15594729 +4403595 to be a hierarchical triple system consisting of a dwarf, SB1 M2, and a companion, M8. The M2 star has a rotation period of P = 0.37\,d, making it the fastest among M-type members of AB Dor. The most relevant result is the detection of a periodic variation in the spotted area on opposite stellar hemispheres, which resembles a sort of Rossby wave or Rieger-like cycles on an extremely short timescale. Another interesting result is the occurrence of a highly significant photometric periodicity, P = 0.443\,d, which  may be related to the stellar rotation in terms of either a Rossby wave or surface differential rotation.} {2MASS J15594729+4403595 may be the prototype of a new class of extremely fast rotating stars exibiting short Rieger-like cycles. We shall further explore what may drive these short-duration cycles and we shall also search for similar stars to allow for a statistical analysis. }
\maketitle

\section{Introduction}
Stellar rotation is a time-dependent quantity that evolves on different timescales during the star's life. In the first stage of life, the evolution of the surface rotation is controlled by the accretion disc through a magnetic locking mechanism (\citealt{Koenigl1991}; \citealt{Shu1994}).
Once the disc is dispersed and the star is free to spin up toward the zero-age main sequence (ZAMS) owing to radius contraction, we deal with a distribution of initial rotation periods for each mass value (\citealt{Gallet2013}; \citealt{Messina2019}). It is only during the main sequence (MS) evolution that the magnetised winds will establish a one-to-one correspondence between mass and rotation period (\citealt{Matt2015}). Such an univocal mass-period relation is first reached by more massive stars (late-F, early-G) and by a few hundred million years it also extends to late-K and early-M stars.
The evolution from ZAMS is the time range in which the gyrochronological method for the age dating of low-mass stars is applied more accurately (\citealt{Barnes2007}; \citealt{Silva-Beyer2022}).\\
\indent
However, close binaries significantly deviate from this scenario,  their surface rotational evolution being influenced by earlier disc dispersal and by the tidal forces between the components. A study by \cite{Messina2019} shows that such tides are effective in altering the rotational evolution with respect to  single stars, when the components are as close as about 100\,au.
These stars play as contaminants of the period-colour distribution and bias the average rotation towards shorter periods, which also means towards younger ages. In this respect, it is particularly challenging to unveil the real nature of very fast rotating stars, which can be either single stars in the convective sequence and migrating towards the interface sequence (\citealt{Barnes2003}) or close binaries whose rotational evolution has followed a different path.
It is mandatory to identify all close binaries in an open cluster, association, or moving group and remove them when the colour-period distribution has to be fitted for dating purposes (see, e.g. \citealt{Messina2017b}, \citeyear{Messina2017a}). 

As part of the RACE-OC project (\citealt{Messina2010}, \citeyear{Messina2011}), we focussed our attention on those members of young open clusters and associations still missing either a spectroscopic or a rotational characterisation. That was the case for 2MASS J15594729+4403595, an M-type likely member of the AB Dor association.
%we focused on the AB Dor association and gathered the rotation periods of all the pre-Gaia members. The study started before both the Gaia and  TESS Data Release and it availed of the photometric ground-based monitoring by different projects, mostly ASAS (\citealt{Pojmanski1998}) and SuperWASP (\citealt{Butters2010}). One specific objective of the project is indeed to carry out ad hoc photometric monitoring to recover the rotation periods missing either from the mentioned surveys or from the literature and to perform  photometric and spectroscopic investigation to identify possible unknown close binaries to prevent them to contaminate and bias the results of the period-color gyrochronological fitting (e.g. \citealt{Messina2014}, \citeyear{Messina2016}).\\
 To measure the rotation period and characterise its possible single or binary nature, we planned a dedicated photometric and spectroscopic monitoring. Our investigation has revealed this star to be particularly interesting owing to its variable behaviour that has never been observed before and that is puzzling in a sense. A summary of literature information is presented in Sect.\,2, our data and literature data in Sect.\,3, periodogram and spectroscopic analyses in Sect.\,4 and 5, the spot modelling results in Sect.\,6, and a discussion and our conclusions in Sect.\, 7 and 8.

 \section{2MASS J15594729+4403595}
2MASS\,J15594729+4403595 (RA = 15:59:47.29; DEC = +44:03:59.5 (J2000); V = 11.86\,mag) is an M2  + M8 visual binary system (\citealt{Lepine2013}; \citealt{Janson2012}). The M8 component was discovered  by \citet{Janson2012} at an angular distance, $\rho$ = 5.638$\pm$0.004${\arcsec}$, and PA = 284.8$\pm$0.3 (epoch 2009.42) from the primary M2 star. \citet{Bowler2015}  confirmed the brown dwarf nature of the secondary M8 component by using near-infrared spectroscopy, inferred a mass of about 43$\pm$9\,M$_J$, and put a limit on the age in the 50-200 Myr range.
The physical association between the two components is inferred by the common proper motion, $\mu$ = 72.7\,mas\,yr$^{-1}$, reported in the NOMAD catalogue (\citealt{Zacharias2005}) 
%The most recent separation measurement is $\rho$ = 5.5$\pm$$0.04{\arcsec}$ (PA = 282.6$\pm$0.3 with Robo-Ao at Kitt Peak with the 2.1 m telescope in late 2016 (Salama et al. 2021).
and confirmed by \citet{Janson2012}, $\mu$ = 73 mas\,yr$^{-1}$. The AstraLux  survey carried out by \citet{Janson2012} allowed the measurement of Sloan magnitude differences of $\Delta$z$^\prime$ = 5.51 mag and  $\Delta i^\prime$ = 6.33 mag between the two components and the following physical parameters to be derived for the primary, A, and secondary, B, components: Sp.T$_{\rm A}$ = M1.0 and  Sp.T$_{\rm B}$ = M8.0, and M$_{\rm A}$ = 0.54 M$_\odot$ and M$_{\rm B}$ = 0.10 M$_\odot$; a separation of 197.3 AU, derived from a spectroscopic distance of 35 pc ($\pm$37\%), which turned out to be underestimated on the basis of the most recent \it Gaia \rm measurements. The latter put this target at a larger distance of d = 42.1$\pm$0.3\,pc and provide a re-normalised unit weight error (RUWE) for the primary component of 13.9, indicating that it is probably an unresolved binary.
\citet{Zickgraf2003} identified 2MASS\,J15594729+4403595 as the optical counterpart of the ROSAT X-ray source 1RXS J155947.5+440358 listed in the RASS-BSC (\citealt{Voges1999}) with Log(L$_{\rm X}$/L$_{\rm bol}$) = $-$3.21. A measure of H$\alpha$ EW =  3.3 \AA\,\,was provided by \citet{Riaz2006} as part of their survey of M dwarfs in the solar neighbourhood.\\
 %Using a new method based on a Bayesian analysis, Malo et al. (2013) find 2MASS\,J15594729+4403595 to be candidate member of the young AB Doradus association with a membership probability of 92.8\% for which the binary hypothesis has a higher probability. For this star they derive a statistical distance d = 33$\pm$4 pc. In subsequent spectroscopic analysis, Malo et al. (2014) measure the radial velocity  RV = $-$15.8$\pm$0.5\,km s$^{-1}$ and the projected rotational velocity $v \sin{i}$ $ = 54.9\pm4.6$\,km s$^{-1}$.
 
%\begin{figure}
%\includegraphics[width=80mm,height=60mm,angle=0]{chart.pdf}
%\caption{Map of a section of the observed field of view.}
%\label{map}
%\end{figure}

\section{Observations}
To investigate the nature of 2MASS\,J15594729+4403595, we carried out photometric and spectroscopic observations and availed ourselves of the observational data that time by time were made available in public archives.
\subsection{Photometry}
 Our own photometric observations were gathered at the Zeta UMa Observatory (Spain) in 2013 (see Appendix\,\ref{app:ZetaUMa} for details on observations and data reduction).\\
About seven years after our ground-based monitoring,  the Transiting Exoplanet Survey Satellite (TESS) started observing 2MASS\,J15594729+4403595, and precisely in Sectors 25, 50, and 51.

\subsection{High-resolution spectroscopy}
Eight high-resolution spectra of  \object{2MASS\,J15594729+440359} were acquired in 2014, in the period between January and July, with the FIbre-fed \'Echelle Spectrograph FIES (\citealt{Frandsen1999}; \citealt{Telting2014}) mounted on the 2.56-m Nordic Optical Telescope (NOT) of Roque de los Muchachos Observatory (La Palma, Spain), under the NOT observing programme P48-217. We used the $1.3\,\arcsec$ medium-resolution fibre, which provides a resolving power of R = 47\,800 and a wavelength coverage of about 3600--7400\,\AA. \\
Following the same observing strategy described in \citet{Gandolfi2013}, in the spectroscopic reduction procedure we started removing cosmic ray hits by splitting each epoch observation into three consecutive sub-exposures and traced the radial velocity (RV) drift of the instrument by acquiring long-exposed (T$_{\mathrm{exp}}$ = 15\,sec) ThAr spectra right before and after each epoch observation. Data were reduced using a customised IDL software suite, which includes bias subtraction, flat-fielding, order tracing and extraction, and wavelength calibration. \\

\section{Periodogram analysis}
\subsection{Zeta UMa Observatory data}
\begin{figure}
\includegraphics[scale=0.37,trim= 10 50 0 100,angle=90]{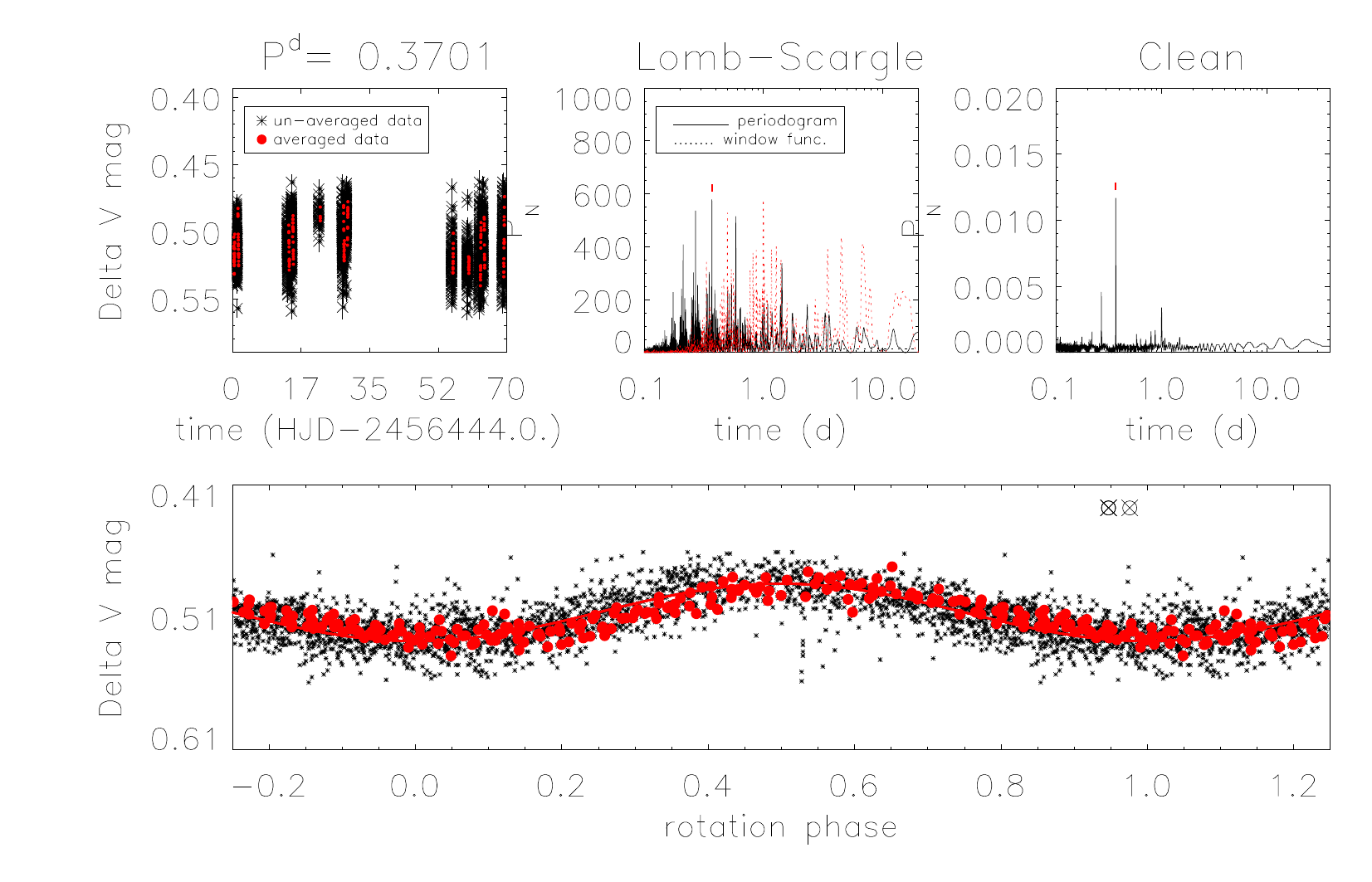}
\caption{Periodogram analysis of Zeta UMa Observatory V-band time series. Top: (from left to right) V magnitude time series; Lomb-Scargle periodogram; Clean periodogram. Bottom: V-band light curve phased with the 0.3701d rotation period. The rotation phase was computed using the following ephemeris: 2456444.0990 + 0.3701$\times$E. }
\label{lightcurve_V}
\end{figure}

\begin{figure}
\includegraphics[scale=0.37,trim= 10 50 0 100, angle=90]{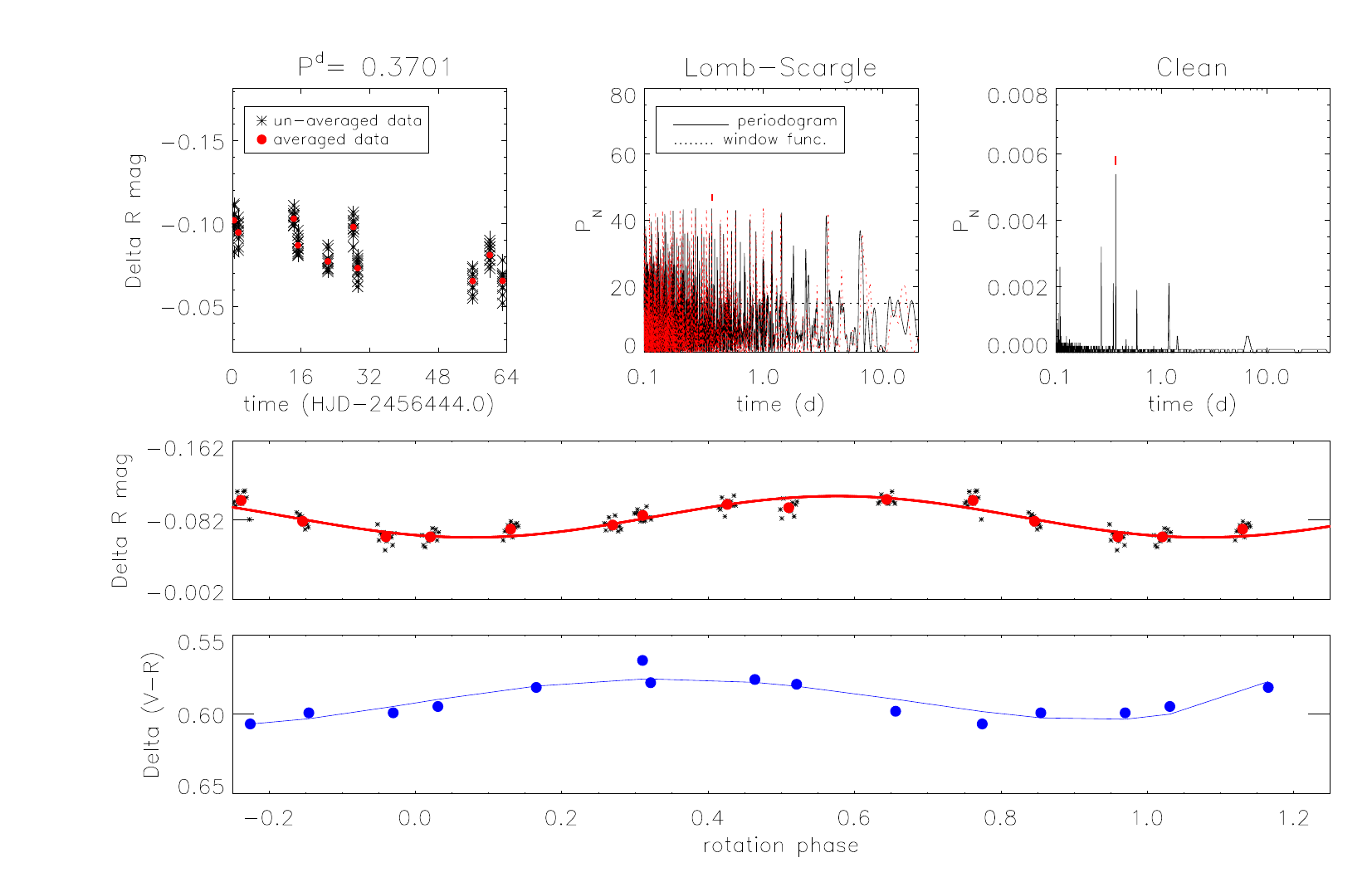}
\caption{Periodogram analysis of Zeta UMa Observatory R-band time series. Top: (from left to right) R magnitude time series; Lomb-Scargle periodogram; Clean periodogram. Middle: R-band light curve phased with the 0.3701d rotation period. Bottom: V$-$R colour curve phased with the 0.3701d rotation period.}
\label{lightcurve_R}
\end{figure}

We used two different periodogram analyses to search for the photometric rotation period in the photometric time series, the generalised Lomb-Scargle (GLS; \citealt{Zechmeister09}), and the CLEAN (\citealt{Roberts87}) periodograms. An estimate of the false alarm probability (FAP) was done using Monte Carlo simulations according to the approach outlined by \citet{Herbst2002}. A more detailed description is given by \citet{Messina2010}. The uncertainty in determining the rotation period was estimated following \citet{Lamm2004} (see also \citealt{Messina2010}). The results of our analysis are plotted in Fig.\,\ref{lightcurve_V} and in Fig.\,\ref{lightcurve_R}.

In the top left panel of Fig.,\ref{lightcurve_V}, we plot the time series of the instrumental differential V magnitudes. Black asterisks and red bullets represent non-averaged and averaged observations (bin width of 15 min), respectively. In the middle panel, we plot the GLS periodogram, with the data window function (dotted red line) overplotted, along with the power corresponding to FAP = 1\% (horizontal dashed line). In the right panel, we plot the CLEAN power spectrum. The GLS periodogram exhibits a number of significant power peaks. The most significant peak is at P = 0.3701$\pm$0.0005\,d, which we assume to be the stellar rotation period. We note that the same period is retrieved by the CLEAN periodogram, which exhibits less peaks owing to its capability to remove the alias arising from the data window function.
In the bottom panel, we plot the light curve phased using the ephemeris 2456444.0990 + 0.3701$\times$E. The V-band phased light curve has an amplitude of $\Delta$V = 0.04\,mag, which was obtained from the amplitude of a fitting sinusoid (solid red line) and exhibits a slight asymmetry, where the phase interval of decreasing brightness is  longer than that of increasing brightness. A similar analysis was carried out with the $\Delta$R magnitude time series. The results are plotted in Fig.\,\ref{lightcurve_R}. 
 The R-band light curve has the same amplitude, $\Delta$R = 0.04\,mag, but unlike the V-band light curve it is symmetric. The differential V$-$R colour curve, plotted in the bottom panel of Fig.\,\ref{lightcurve_R}, has an amplitude of $\Delta$(V$-$R) = 0.03\,mag. The colour curve has its maximum around phase $\phi$ $\simeq$ 0.3, whereas the V- and R-band light curves have their maximum around phase $\phi$ $\simeq$  0.5.\\
In the case of variability dominated by cool spots, we generally observe a positive correlation between the light and colour curve; when spots are most visible, and therefore the light curve has a minimum, the star is reddest, and the colour curve exhibits a minimum.  In the case of variability dominated by cool and hot spots, we generally observe a negative correlation between the light and colour curves; when spots are most visible and the light curve has a minimum, the star is bluest, and the colour curve exhibits a maximum (see, e.g. \citealt{Messina2008}). In the case of  
2MASS J15594729+4403595, we observe during the 2013 season a sort of mixed case in which two-temperature inhomogeneities drive the observed variability; however, they are not at the same mean stellar longitudes,  their photometric barycenters being separated by $\Delta \phi \simeq$ 0.2, or about 70 degrees, giving rise to the aforementioned asymmetries.\\
To infer the inclination of the stellar rotation axis, we combined the rotation period, stellar radius, and projected rotational velocity. The stellar radius was derived from the evolutionary track best fitting our target in the HR diagram.
The 2MASS and \it Gaia \rm magnitudes of the primary component (resolved from the M8 component in both photometries) provide colour indices in agreement with a M2 spectral type. Taking into account a slight bias towards redder wavelengths owing to the unresolved companion of the primary (SB1) component, we can assume for it a spectral type of M1.5$\pm$0.5 and an effective temperature of T$_{\rm eff}$ =  3500\rm $\pm$50\,K according to the \citet{Pecaut2013} relations.
In the HR diagram, the primary component of the SB1 system is best fit, after correction for the magnitude contribution of the unresolved unseen component, by an evolutionary track of mass 0.45$\pm$0.05\,M$_\odot$ and an age in the range of 30--80\,Myr (\citet{Baraffe2015}). The corresponding stellar radius is R$_{\odot}$ = 0.50$\pm$0.05. Combining the radius, rotation period, and projected rotational velocity (see Sect.\,5), we derive an inclination for the stellar rotation axis of $i$  = 50$\pm$10$^\circ$.

\subsection{TESS data\label{TESS}}
\begin{figure}
\label{periodogram_TESS_Sect25}
    \includegraphics[scale = 0.3, angle = 90]{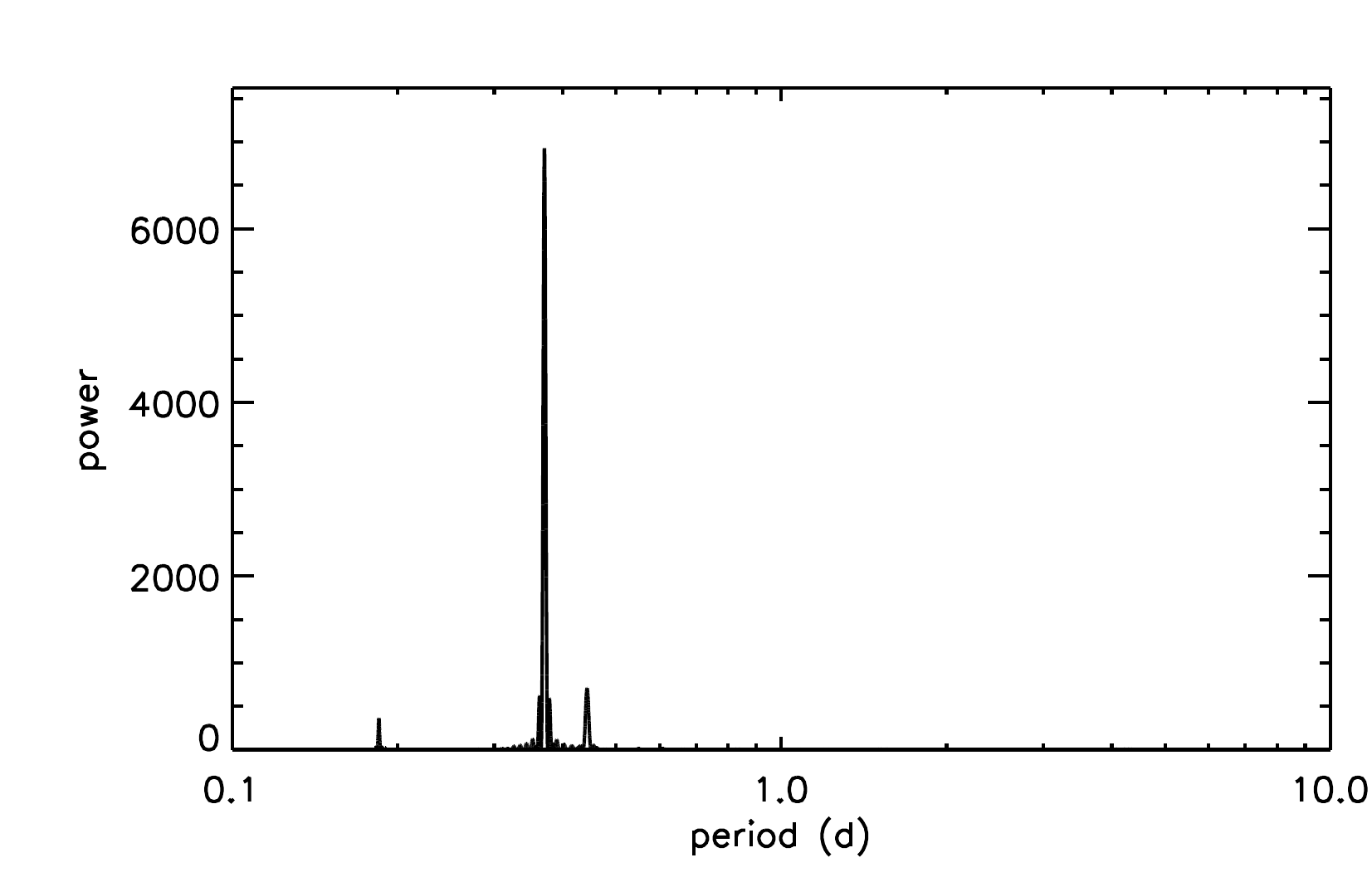}\\
      \caption{Periodogram of TESS time series collected in Sect.\, 25. The power peaks corresponding to the three periodicities, P$_1$ = 0.370\,d, P$_2$ = 0.443\,d, and  P$_3$ = 0.18498\,d, are clearly detected. Similar periodograms are obtained when analysing Sectors 50 and 51.}
\end{figure}

\begin{table*}
\caption{Results of periodogram analysis on TESS data.}
\label{TESS_table}
\begin{tabular}{l|c|c|c|c|c|c}
\hline
 TESS & P$_1$ & P$_2$ & P$_3$ & A$_1$ & A$_2$ & A$_3$ \\
 Sector & (d) &(d) &(d) & (mag) & (mag) & (mag)\\
  \hline
Sector 25 & 0.370$\pm$0.003 & 0.443$\pm$0.004 & 0.1849$\pm$0.0006 & 0.016 & 0.005 & 0.004\\
Sector 50 & 0.370$\pm$0.003 & 0.443$\pm$0.004 & 0.1849$\pm$0.0006 & 0.015 & 0.006 & 0.003\\
Sector 51 & 0.370$\pm$0.003 & 0.443$\pm$0.004 & 0.1849$\pm$0.0007 &
0.018 & 0.007 & 0.002\\Sector 25-51 & 0.3700$\pm$0.0009 & 0.4429$\pm$0.0001 &
0.1850$\pm$0.0002 &
0.016 & 0.006 & 0.0036\\
\hline
\end{tabular}
\end{table*}
\begin{figure}
\label{TESS_figs}
    \includegraphics[scale = 0.3, angle = 90]{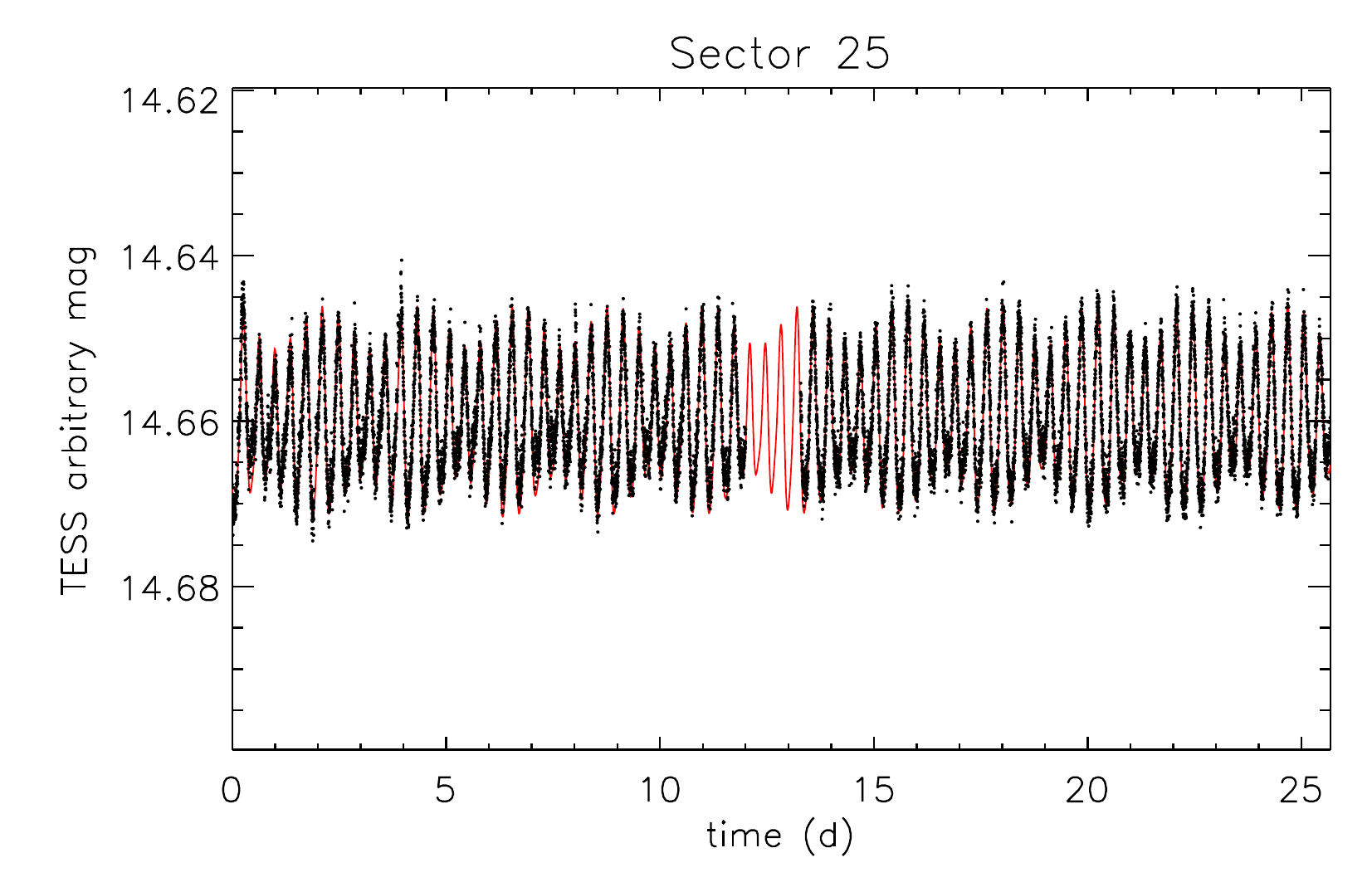}\\
     \includegraphics[scale = 0.3,angle = 90]{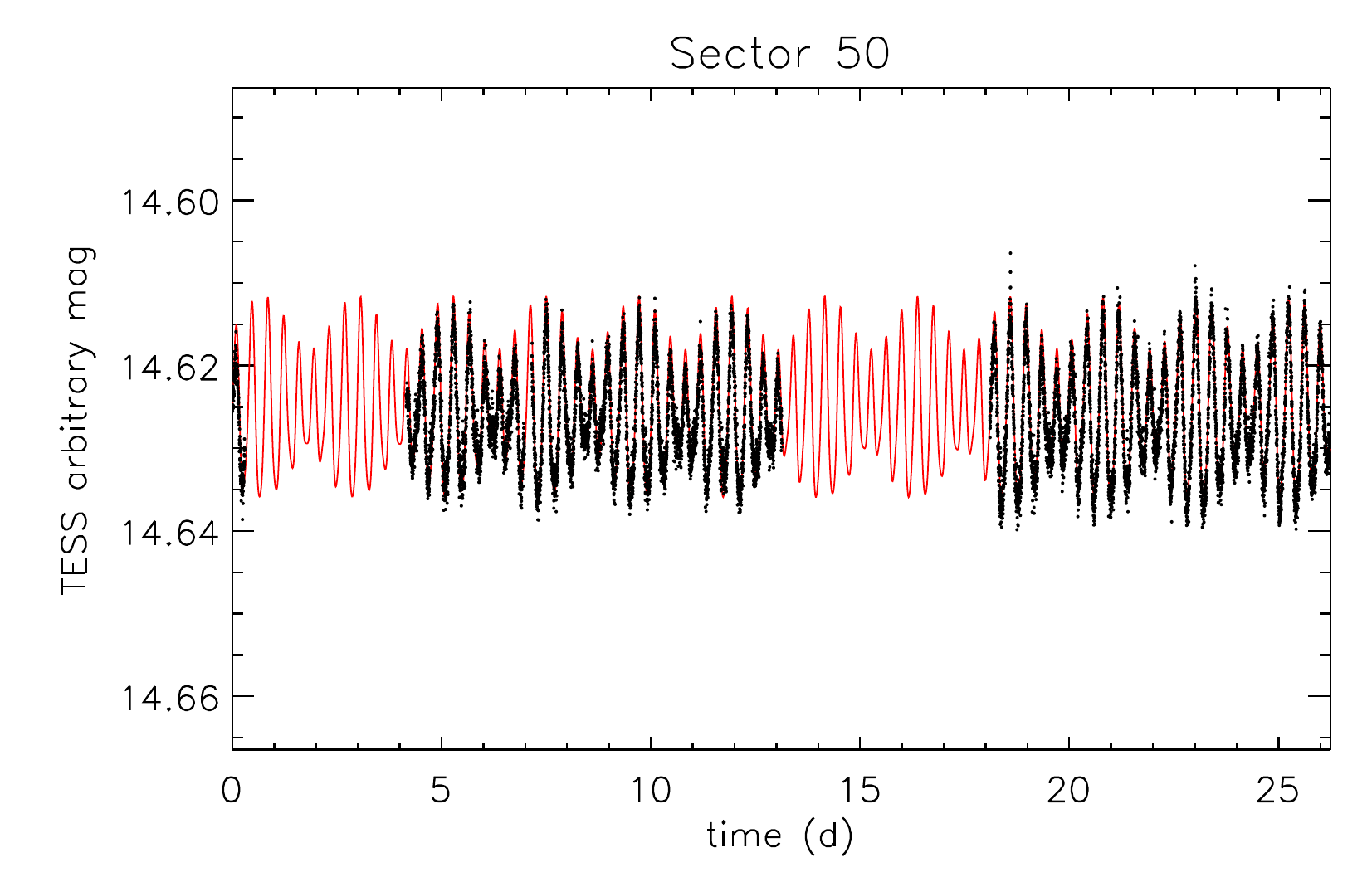}\\
      \includegraphics[scale = 0.3,angle = 90]{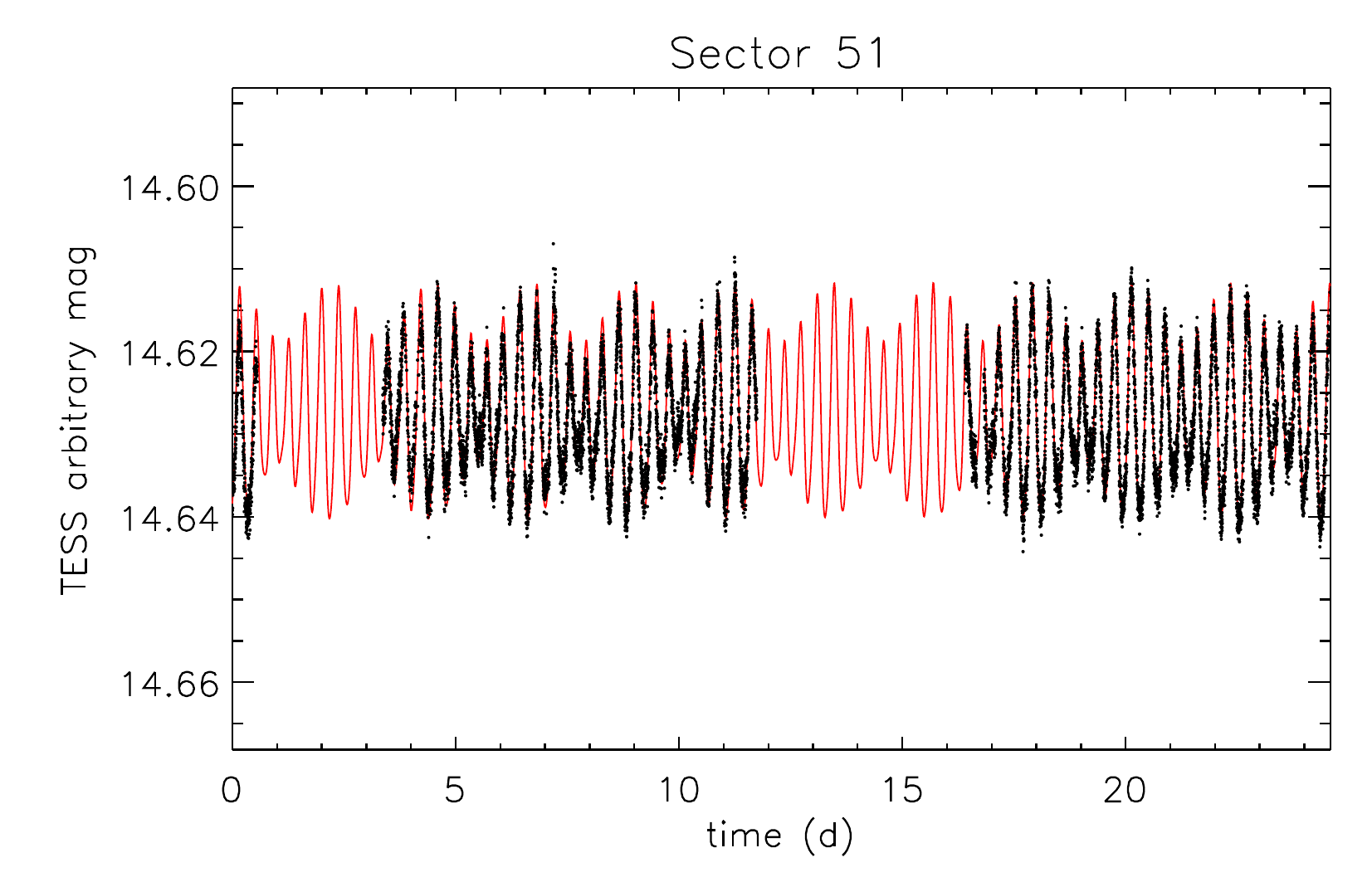}\\
      \caption{TESS time series (black dots) from Sectors 25, 50, and 51 with overplotted sinusoidal fits (red line) with the three periodicities P$_1$ = 0.36996\,d, P$_2$ = 0.44298\,d, and  P$_3$ = 0.18498\,d.}
\end{figure}
2MASS\,J15594729+4403595 was observed by TESS in Sectors 25, 50, and 51.   In the following periodogram analyses, we used the PDCSAP fluxes extracted from the Mikulski Archive for Space
Telescopes (MAST). The results of our periodogram analysis are listed in Table\,\ref{TESS_table}.\\
In each sector and in their combination, we detected three highly significant periodicities, P$_1$ = 0.36996 $\pm$0.00009\,d, P$_2$ = 0.44298 $\pm$0.0001\,d, and
P$_3$ = 0.18498$\pm$0.00002\,d (see Fig.\,\ref{periodogram_TESS_Sect25} as an example). We first note that P$_1$ is in very good agreement with the period, P = 0.3701$\pm$0.0005\,d, measured from the Zeta UMa data. Then, we note that  P$_3$ = P$_1$/2. The more plausible explanation for the presence of P$_3$ is that it is the first harmonic of P$_1$ and it arises from the presence of spots on opposite hemispheres of the stellar surface, which are
carried in and out of view with a period, P$_1$, and produce a double-dip modulation with minima
of unequal depth,  their respective areas being different.
Indeed, the amplitude, A$_3$, of the rotational modulation associated with P$_3$ is about 25\% of A$_1$. The presence of P$_2$ is challenging (see Sect.\,\ref{Discussion}). 
In Fig.\,\ref{TESS_figs}, we plot the photometric time series corresponding to the three TESS Sectors with, overplotted, the sinusoidal fit (solid red line) with all three periodicities. It is clearly visible the beating between period P$_1$ and P$_2$ which, due to their difference of approximately 15\% produces a modulation of the variability amplitude with a period of about P$_1$/0.15 = 2.5 days.

\section{Physical and orbital parameters}

We estimated the projected rotational velocity, $v \sin{i}$ of 2MASS\,J15594729+440359, by using the spectrum of GJ\,411 as a template. The spectral lines of GJ\,411 are unresolved in the FIES spectra, owing to the small $v \sin{i}$ of the template.\footnote{$v\sin{i}$  = 1.6\,km s$^{-1}$ according to \citet{Glebocki2005}.} Assuming a linear limb-darkening coefficient, $\mu = 0.65$  (for the TESS waveband centred at 786.5 nm), \rm for early-type M-dwarf stars (\citealt{Claret2012}), we found that the projected rotational velocity of 2MASS\,J15594729+440359 is $v \sin{i}$ $ = 54\pm5$\,km s$^{-1}$.

%Radial velocity measurements were derived performing a multi-order cross-correlation with the early type M-dwarf star \object{GJ\,411} -  observed with the same instrument set-up as the target object -  and for which we adopted a heliocentric RV~$=-84.689$ km s$^{-1}$ as measured by Nidever et al. (2002).\\

Radial velocity measurements were derived by performing a multi-order cross-correlation with a synthetic template computed with the atmospheric parameters typical of an M2V star and broadened for the resolving power and rotational velocity of our target. The cross-correlation was calculated paying particular attention to excluding Balmer lines from the correlation, as well as spectral intervals with telluric lines. The FIES RV measurements are listed in Table\,\ref{RV-Table}, along with the heliocentric Julian date (HJD) of mid-exposure, the total exposure time, and the S/N per pixel at 5500\,\AA. 

The FIES spectra reveal a rapidly rotating star and strong emission in the Balmer and Ca\,{\sc ii} H and K lines, confirming the high magnetic activity level suggested by the strong X-ray luminosity. The cross-correlation function of the various epoch spectra shows an asymmetric and variable profile that we ascribe to magnetically active regions carried around by stellar rotation. 

{ As the reader can easily see in Table\,\ref{RV-Table}}, our RV measurements are not consistent with each other. This difference may arise from orbital motion and suggests the existence of a fainter close-in companion. Therefore, the bright M2 component may be an SB1 spectroscopic binary. To investigate this hypothesis, we searched the literature for other RVs and several spectroscopic archives. The RVs of our target have been published by \citet{malo14}, \citet{binks16}, and \citet{jon2020}.
Moreover, we downloaded three spectra from the CFHT archive acquired in 2012, 2013, and 2014 with ESPADONS. We derived RVs for these spectra with the same procedure as we did for FIES data. In particular, the 2013 spectra is the one published by \citet{malo14}, so we used this spectrum to check the validity of our methods. Our determination of RV is totally compatible with that published by the latter authors.

%Our RV measurements of $-$27.5 and $-$28.6 km s$^{-1}$- whose error bars are strongly affected by the high rotation rate of the star and the distorted CCF profile - are consistent within $\sim$2$\sigma$, and are significantly different from the value measured by \citet{malo14}. This difference may arise from orbital motion and suggest the existence of a fainter close-in companion. Therefore, the bright M1 component may be an SB1 spectroscopic binary. 

A GLS periodogram analysis (\citealt{Zechmeister09}) of the RV time series did not reveal any significant periodicity. Therefore, in order to get a tentative orbital solution for our system, we proceeded as  follows.

The RVs for a spectroscopic binary system are given by the following equation:

\begin{equation}
\label{radteo}
V_{\rm rad} = \gamma + K [\cos(\theta + \omega) + e \cos \omega]
,\end{equation} 

\noindent where $\gamma$ is the RV of the centre of mass,
$e$ is the eccentricity of the orbit, $\omega$ is the longitude of the periastron,
$\theta$ is the true anomaly of the orbital motion at 
a given instant, and $K$ is the semi-amplitude of the velocity curve given by the
formula
 
\begin{equation}
K = \frac{2\pi a\sin i}{P\,\sqrt{1-e^2}}
,\end{equation}

\noindent where $P$ is the orbital period of the system, $a$ its semi-major axis, and $i$ the 
inclination angle.

By using all the velocities, orbital elements were determined by a weighted least-squares fitting to Eq.~\ref{radteo}. Errors were estimated as the variation in the parameters that increases the $\chi ^2$ of a unity. 

\begin{table}
  \caption{FIES/NOT and ESPADONS/CFHT  RV measurements. }
  \label{RV-Table}
\begin{tabular}{ccccc}
\hline
\noalign{\smallskip}
HJD              &   RV    & $\sigma_{\mathrm RV}$ &   T$_{\mathrm{exp}}$  &  S/N/pixel    \\
&   km s$^{-1}$ &    km s$^{-1}$               &       s               &  @5500\,\AA   \\
\noalign{\smallskip}
\hline
\noalign{\smallskip}
%2456665.76941417 & $-$27.53  &         0.49        &  2100   &   31  \\
%2456668.74116662 & $-$28.66  &         0.34        &  3600   &   49  \\
\noalign{\smallskip}
\multicolumn{1}{c}{FIES/NOT}  & & & &\\
56665.76941 & $-$28.3 &   0.3  & 2269 & 31 \\ 
56668.74116 & $-$29.6 &   0.3  & 3769 & 49 \\ 
56739.72624 & $-$21.7 &   0.3  & 2872 & 60 \\
56740.68460 & $-$20.8 &   0.3  & 5742 & 70 \\ 
56741.68671 & $-$22.1 &   0.3  & 2869 & 55 \\ 
56841.39334 & $-$13.5 &   0.3  & 1887 & 60 \\ 
56842.39381 & $-$13.7 &   0.3  & 1887 & 50 \\ 
56855.41977 & $-$13.8 &   0.3  & 1287 & 40 \\ 
\hline
\noalign{\smallskip}
\multicolumn{2}{c}{ESPADONS/CFHT}   & & &\\
55970.11628 & $-$23.5 &   0.5  & ~~500 & 30 \\      
56355.12240 & $-$15.5 &   0.4  & 1000 & 30 \\                      
56825.99784 & $-$16.2 &   0.3  & 1200 & 30 \\ 
\noalign{\smallskip}
\hline
\end{tabular}
\tablefoot{The total exposure time and the S/N per pixel at 5500\,\AA\ are listed in the last two columns.}
\end{table}

\begin{tabular}{ll}
 & \\
$P$:&  13.976 $\pm$ 0.001 d \\
$T$:& 2457532.966 $\pm$ 0.060\\
$e$:& 0.35 $\pm$ 0.03 \\
$\omega$:& 148$^\circ$ $\pm$ 2$^\circ$ \\
$K$:& 11.0 $\pm$ 0.5 km s$^{-1}$\\
$\gamma$:& $-$20.5 $\pm$ 0.2 km s$^{-1}$\\
$a \sin i$:& 2.8 $\pm$ 0.2 R$_\odot$\\
$f(m)$:& (1.6 $\pm$ 0.3) $\cdotp 10^{-3}$ M$_{\odot}$\\
  & \\
\end{tabular}

\noindent where the mass function, $f(m)$, is defined as

\begin{equation}
 f(m) = \big(\frac{M_2}{M_1+M_2}\big)^2 M_2 \sin^3 i 
\label{fm}
.\end{equation}

\begin{figure}
\includegraphics[width=\columnwidth]{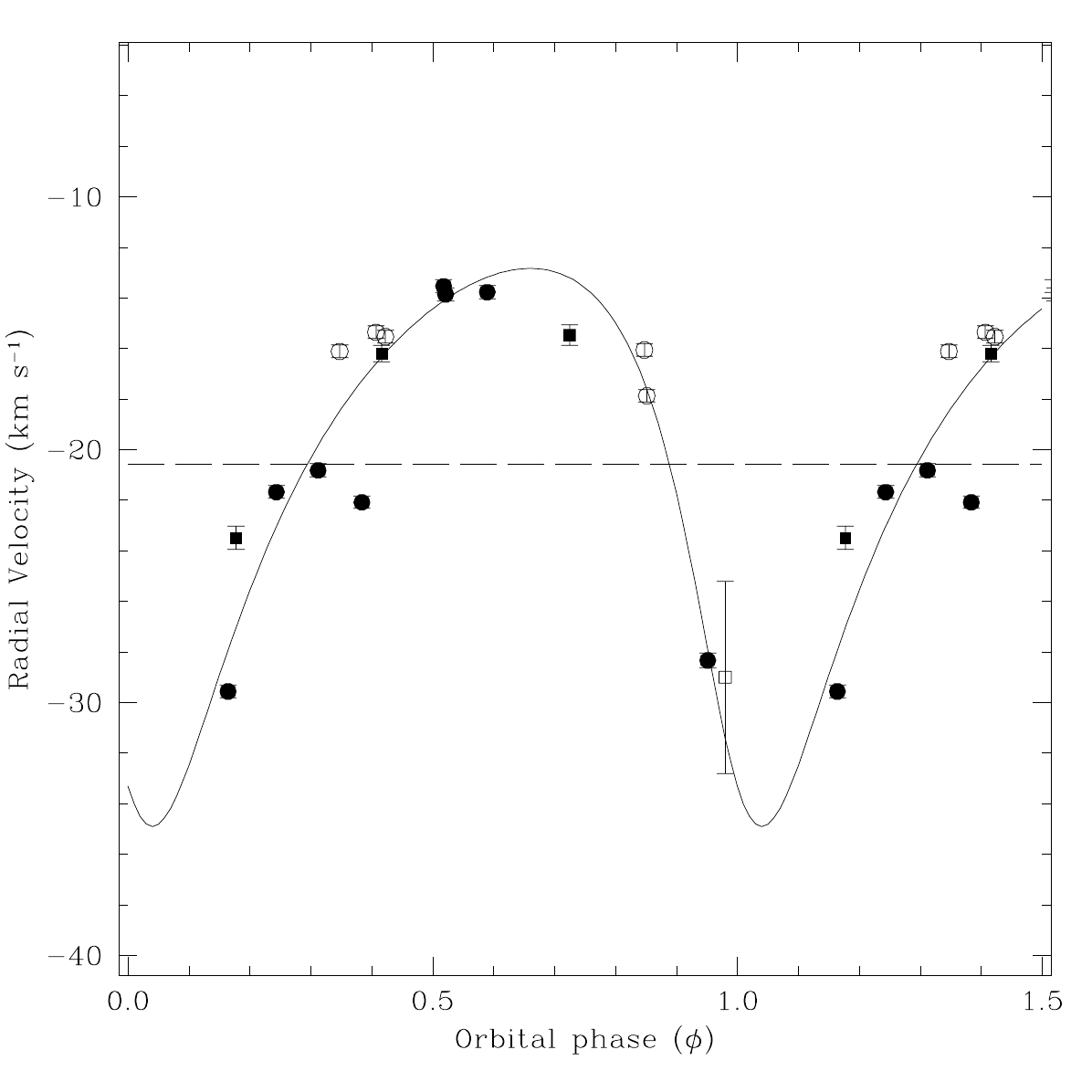}
\caption{Radial velocity curve for our target. The black dots represent RVs from FIES/NOT and the black squares ones from ESPADONS/CFHT. The error bars are as large as three $\sigma$. The literature value from \citet{binks16} is plotted as an open square, and those from \citet{jon2020} as open circles. }
\label{orbit}
\end{figure}

This solution has been overplotted on the measured velocities in Fig.~\ref{orbit}.

\begin{figure}
\includegraphics[width=\columnwidth]{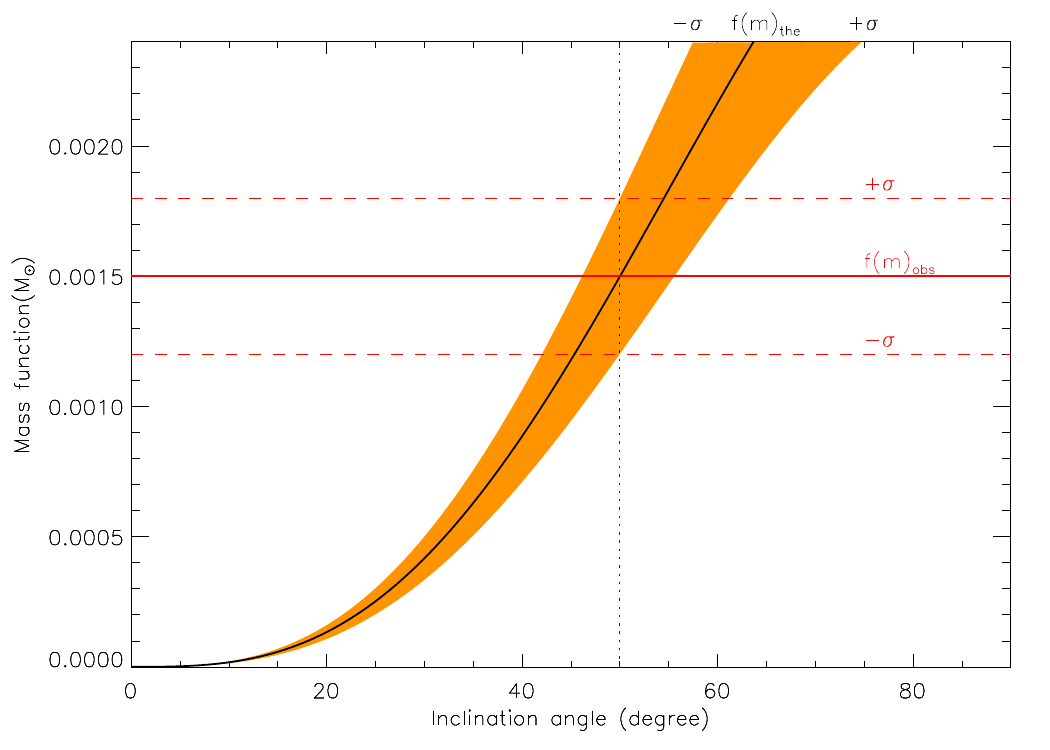}
\caption{Theoretical mass function vs the inclination angle. Horizontal red lines represent the mass function derived from the orbital solution. The vertical blue dotted line represents the inclination angle estimated for our star. The orange surface represents the $\pm 1\,\sigma$ confidence level. The inclination angle of i = 50$^\circ$ (see Sect.\,4.1) implies a theoretical mass function computed with M$_2$ (M$_\odot$)\,=\,0.0245\,$^{-0.0027} _{+0.0025}$.}
\label{funcmass}
\end{figure}

By using the previous solution, we have speculated about the mass of the companion under two hypotheses on the mass of the primary, which is fixed to M$_1$\,=\,0.40\,M$_\odot$ \citep{allen}, and on the inclination angle of the orbit, $i$ = 50$^\circ$ (see Sect.\,4.1). Solving Eq.~\ref{fm} graphically, as is shown in Fig.~\ref{funcmass}, we obtained a value for the companion of M$_2$\,(\,M$_\odot$) = \,0.0245\,$^{-0.0027} _{+0.0025}$. We note that since the orbital period, P$_{orb}$= 13.976\,d, is not supported, even if not disproven, by the periodogram analysis  of the RV time series, \rm we conservatively consider our orbital solution to be tentative.

\begin{figure*}

    \includegraphics[scale = 0.70, angle = 0, trim = 30 0 1200 0]{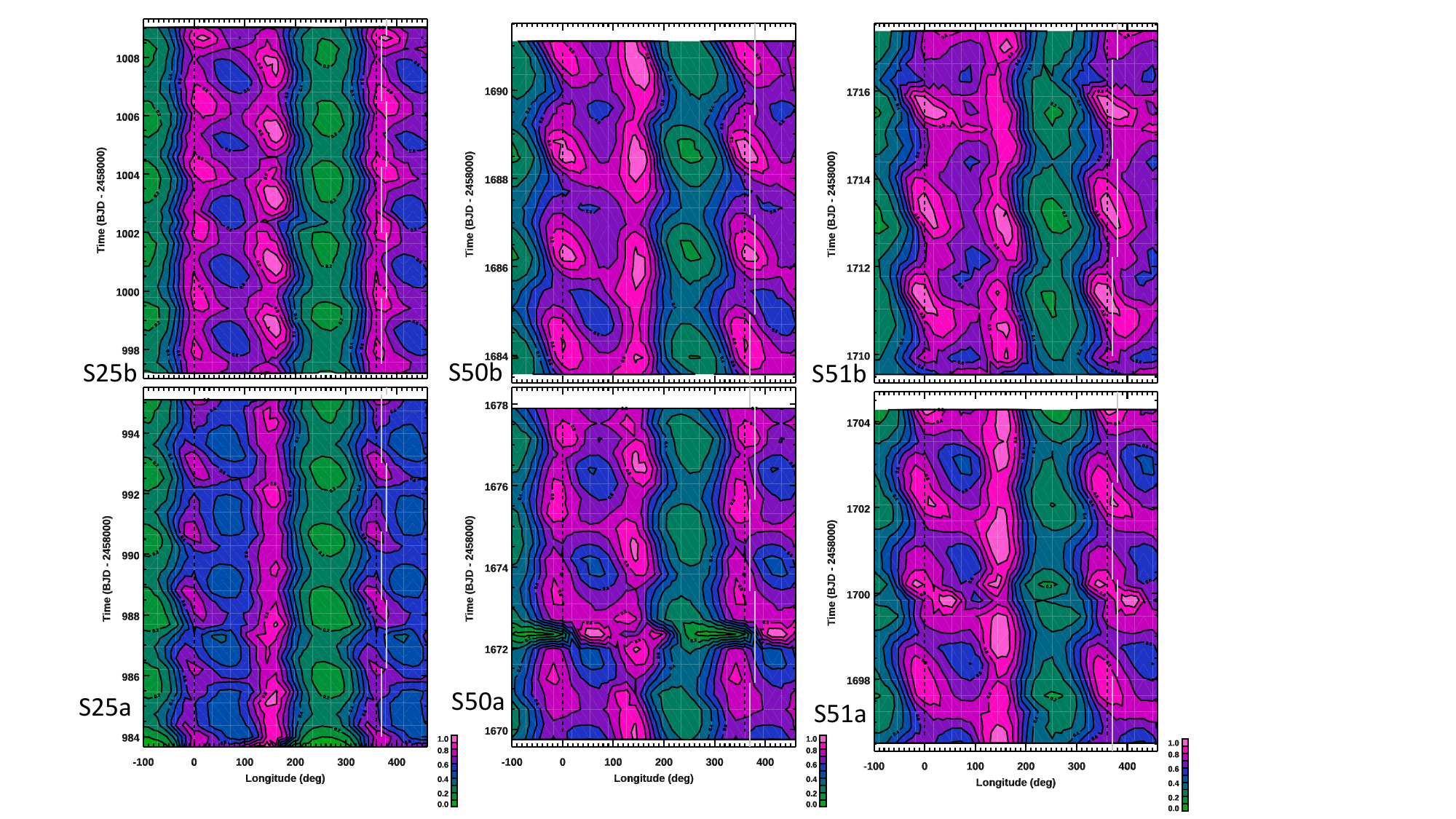}
    \caption{\label{maps}Distribution of the spot filling factor vs the longitude and time derived by
our maximum-entropy spot model of the TESS light curves in Sectors 20, 50, and 51. The maximum of
the filling factor is indicated by the purple colour and the minimum by dark blue
(see colour scale in the lower right corners).
We note that the longitude scale is repeated
beyond the [0$^\circ$
, 360$^\circ$
] interval to better follow the migration of the spot features. 
Each TESS sector is split into two intervals to skip the data gap. The solid white lines help one to track the spot evolution from one active longitude to the other.}
\end{figure*}

\section{Spot model}
\label{sec_spot_model}

As will be discussed in Sect.\,\ref{Discussion}, the observed photometric variability among different mechanisms may originate from surface brightness inhomogeneities of a magnetic nature. \rm
In this case, in order to investigate the surface distribution of brightness inhomogeneities (spots) on 2MASS J15594729+4403595 and their time evolution, we applied the spot modelling approach already introduced in Sect. 3 of \citet{Bonomo2012}, to which we refer the reader for details. 
A brief summary is given in Appendix \ref{spot_model}.
As a result of our modelling, we obtained a time series of maps of the spotted surface of 2MASS J15594729+4403595. 
The maps versus time of the spot distribution across the photosphere are plotted in Fig.\,\ref{maps}. From left to right, the maps refer to  the TESS Sectors 25, 50, and 51. In each sector, time series have been divided into two intervals to prevent the data gaps (see Fig.\,3) from affecting the modelling. 

We note that spot activity in our reference frame is permanently located at two longitudes centred at about 0$^\circ$ (hereafter, LongA) and 160$^\circ$ (hereafter, LongB), whereas a very marginal activity level is at 270$^\circ$ (hereafter LongC). We selected the origin of the longitudes so as to have the first active longitude at longitude zero at the beginning of our time series.
On longitudes LongA and LongB, the level of spottedness evolves periodically with time, with a period of about 2.25 days, resulting from a periodogram analysis of their spotted area. However, the activity levels in the two longitudes are approximately in anti-phase; that is, when the activity is maximal at LongA, it is minimal at LongB and vice versa. Handmade vertical white lines help us to track the periodicity of 2.25 days. Therefore, we observe two brightness waves on opposite hemispheres that evolve in about six stellar rotations and that are in anti-phase.

\begin{figure}
    \includegraphics[scale = 0.35, angle = 90, trim = 0 0 0 0]{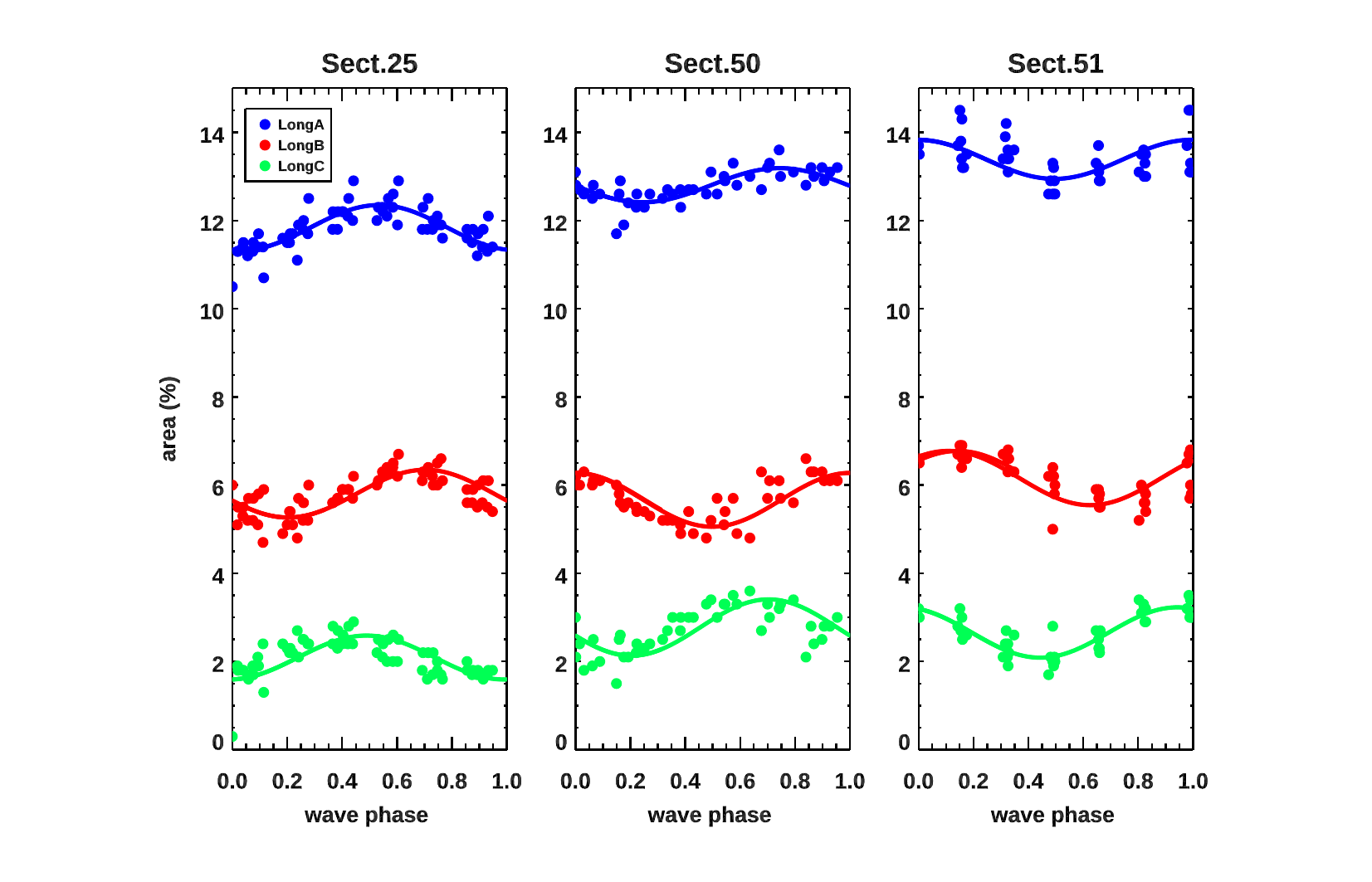}
    \caption{\label{spot_area} Spot area at each active longitude vs the phase of the presumed Rieger cycle, whereby phases were computed using the area modulation period, P = 2.25\,d. It is evident that spots A and B share the highest activity level, whereas spot C has marginal activity. The area modulation of spots A and B is in opposition, with activity at maximum in one longitude when at minimum in the opposite longitude.}
\end{figure}

The periodic spottedness change at each active longitude is more visible when the respective spot area is plotted versus its cycle phase, as in Fig.\,\ref{spot_area}.\\ In principle, the retrieved spot pattern may be  an artifact of the model, arising from the attempt of the spot code to reproduce the beat-like pattern of the observed light curve. To  explore \rm this possibility, we performed two tests. They consist of hypothesising \rm two different causes for the observed variability (specifically, surface differential rotation (SDR) and Rieger cycles), generating synthetic light curves, and checking whether the maps are consistent with the hypothesised scenarios. Both tests are discussed in more detail in Appendix\,\ref{spot_model}. 
As a result, comparing our Tests 1 and 2 with the maps obtained from the observed data, we are confident that the maps we retrieved from real data are not an artifact but consistently reproduce a physical phenomenon.
\rm

\section{Discussion}
\label{Discussion}

The RV variability unveiled by our measurements complemented with the ones from the literature and the large RUWE value of about 13 reported in the \it Gaia \rm DR3 allow us to state that the primary component of 2MASS\,J15594729+4403595 is a close binary and, since no evidence of spectral features of its secondary have been detected in our high-resolution spectra, we can classify it as an SB1 system. This means that 2MASS\,J15594729+4403595 is a hierarchical triple system, consisting of an unresolved M2+? close binary and a visual M8 component. Unfortunately, neither TESS nor Zeta UMa observations could spatially resolve the two components of the SB1 system or the primary M1 from the secondary M8 component, their spatial resolutions being  21$^{\arcsec}$/pixel and 1.5$^{\arcsec}$/pixel, \rm respectively.
Since 2MASS\,J15594729+4403595 is a photometrically unresolved system in our study, the analysed photometry contains signals from all three components. However, the M8 component is about six magnitudes fainter than the primary in the $i^{\prime}$ band, which implies that its flux contribution is about a factor of 250 smaller than that of the primary. Therefore, its contribution to the observed variability is negligible.
Since the secondary unseen component of the SB1 is expected to be much fainter than the primary, as our spectroscopic analysis indicates, its contribution  to the total flux is also expected to be negligible. In this circumstance,  the observed photometric variability is dominated by the primary component.\\
If this is the correct scenario, we deal with a star with a rotation period of P$_1$ = 0.3700\,d and spots on opposite hemispheres that produce a secondary rotational modulation with a period given by the first harmonic (P$_3$ = 0.1850\,d). \\
\indent
Interpreting the presence of the periodicity, P$_2$ = 0.44298\,d, is challenging. 
In this context, it is relevant to remind ourselves that the P$_2$ periodicity clearly detected during TESS observations was not detected in the data time series collected at the Zeta UMa Observatory in 2014, although that photometry was precise enough to detect it, if it were present. In this respect, it will be very interesting to check if P$_{\rm 2}$ is also absent in the \it Gaia \rm time series that was collected almost contemporaneously, but that has not yet been released.
In what follows,  we present different possible interpretations for the existence of the P$_2$ = 0.44298\,d periodicity.
\indent
\subsection{Ellipsoidicity}
A first hypothesis to account for the P$_{2}$ periodicity is that the primary component has the  secondary so close as to suffer from some level of ellipsoidal tidal deformation. In such a case, the changing projected surface of the primary can produce a rotational modulation with the SB1 orbital period of P$_2$ = 0.4429\,d. In this case,  we were dealing with an un-synchronised system for which the axial rotation period, P$_1$, is smaller than the tentatively estimated orbital period, P$_{\rm 2}$. However, our periodogram analysis  of the RV time series \rm does not show any significant power peak at such a period, whereas the best orbital solution has an orbital period of about 14 days. Moreover, in this hypothesis, the periodicity should be permanent, whereas, as was mentioned, it was not detected in the data of the 2013 photometric campaign. 
We could consider also the case of a star that is precessing without having a disc. The precession would be caused by a misalignment between the stellar equator and the orbital plane, which could be possible, given that the star is relatively young. Nevertheless, this model implies that the periodicity is permanent, which is not in agreement with the observations.  Moreover, we would expect a precession period longer than the orbital period; that is, longer than 14 days. 
Were the \it Gaia \rm photometry to confirm that P$_{\rm 2}$ was not present, we could exclude the ellipsoidicity hypothesis and  consider the following hypotheses.  \rm

\subsection{Clumpy disc}
Another possible interpretation to account for the $P_{2}$ periodicity is that 2MASS J15594729+4403595 has a disc with some sort of clumpy structure that, when the precession allows the disc to pass in front of the star, periodically transits across the stellar disc (see, e.g. \citealt{Rodriguez-Ledesma2012}) and produces the observed rotational modulation with a keplerian period, P$_2$, longer than the stellar rotation period, P$_1$.  However, 
%the low value of the inclination ($i$ = 50$^{\circ}$) makes unlikely our hypothesis that a dust clump close to the star's surface may transit across the disc and may produce the observed rotational modulation. On the other hand, also 
the estimated age does not fit well with the presence of a dusty disc  \citep[e.g. ][]{Ribas2015}, unless it is a debris disc. \\

\subsection{Surface differential rotation}

 Another  possible interpretation \rm is in terms of SDR. In fact, on a differentially rotating star, spots at different latitudes can produce rotational modulations of different periods. If this is the case, the relative rotation period difference, (P$_2$$-$P$_1$)/P$_2$, would imply an SDR amplitude of about 16\%. According to \citet{Kueker2019}, such a star should have a differential rotation not larger than a few percent owing to its very fast rotation. Very fast rotators are expected to behave as rigid bodies. In a sample of over 18000 Kepler stars with measured SDR, \citet{Reinhold2013} found the relative shear, $\Delta\Omega$/$\Omega$, to decrease with a decreasing rotation period, and to further slightly decrease with effective temperature. \rm
On the other hand, according to the simulations by \citet{Brun2022}, this large value is not unlikely. Depending on the latitudes at which major spot centres are located,  differential rotation may not be detectable, as it happened in the 2013 observations.\\
\indent

\subsection{Rossby waves and Rieger cycles}

We considered the frequency, $\sigma_{\rm obs}$, of a Rossby wave seen by a distant observer,
\begin{equation}
\sigma_{\rm obs} = m\Omega + \sigma, 
\end{equation} 
with the frequency of the wave in a reference frame rotating with the star  given by 
\begin{equation}
    \sigma = -\frac{2m \Omega}{n(n+1)},
    \label{rossby_freq}
\end{equation}
where the minus sign comes from the retrograde propagation of the wave in the rotating frame, $m$ is the azimuthal wavenumber, $\Omega \equiv 2\pi/P_{\rm rot}$ the stellar rotation frequency, and $n$ the degree of the wave \citep[see][Eq.~30 for details]{Zaqarashvilietal21}. The transformation of the frequency from the reference frame rotating with the star to that of a distant observer is discussed in, for example, \citet{Kepler84} and \citet{Unnoetal89}. In principle, the effects of the rapid rotation and strong magnetic fields on the eigenfrequencies of the Rossby waves  cannot be neglected. An analysis is presented in Appendix~\ref{app_rossby}, in which we show that the centrifugal deformation of the star has a negligible effect, provided that the wave is propagating close to the interface between the convection zone and the radiative interior, while strong magnetic fields can be neglected because the Coriolis force is much stronger than the Lorentz force in our rapidly rotating star.

In the specific case of the active component of 2MASS\,J15594729+4403595, we adopted $n=3, m=1$, which give a periodicity in the observer's frame of $(6/5)P_{\rm rot}= 0.444$\,d, when a rotation period, $P_{\rm rot}=P_{1}=0.370$\,d, is adopted. Rossby waves are waves of vorticity; therefore, they do not directly produce  brightness oscillations as in the case, for example, of p-mode waves, which produce temperature and density perturbations. Nevertheless, considering that those waves can affect the emergence and the distribution of magnetic structures — for example, in the solar corona \citep[cf. Sect.~4.2 of][]{Zaqarashvilietal21} — it is not excluded that they can produce brightness oscillations in our very active star, thereby accounting for the $P_{2}$ periodicity as seen by a distant observer. On the other hand, a modulation of the magnetic flux emergence possibly associated with the same Rossby wave has a period of $2\pi/\sigma = 2.25$\,d in the reference frame rotating with the star. Such a modulation can induce a short-term activity cycle akin to a solar Rieger cycle in our star, as is discussed by \citet{Zaqarashvilietal21} in their Sects.~4.2.4-4.2.6. 

To further discriminate between the SDR and the Rieger cycle  hypotheses, the spot model comes into help. As part of Test 1 in Appendix\,\ref{spot_model},  we built one synthetic light curve, as was produced in the case of SDR, with two active regions rotating with P$_1$ and P$_2$ periods, and ran our code to produce the corresponding map time series. Consistently \rm with the proposed scenario, we find the total spot area (summed over all longitudes) to stay constant during consecutive rotation cycles. Similarly, as part of Test 2, we built one synthetic light curve, as was produced in the case of a cyclic modulation of the spot area induced by a Rossby wave with a period of P = 2.25\,d in the stellar frame. We also added the brightness change produced by the wave as well as the rotational modulation of the visibility of the active regions on the stellar surface as they are seen by a distant observer. 
A comparison between the SDR-case and Rieger-case spot maps with those retrieved from the observed TESS time series indicates the Rieger cycle hypothesis to be the more likely one. %The rotation period P$_1$ and the periodic brightness variation P = 2.25\,d combines in the synthetic time series in such a way as to account for the observed P$_2$ periodicity and the corresponding maps show a pattern similar to that inferred from real data.[Questa parte è un ripetizione di quanto detto prima e conviene toglierla per non conforndere il lettore.]

\rm

%Unless the disc is misaligned with respect to the equatorial plane, this hypothesis requires that the star be seen almost equator on.
%\begin{figure}
%    \includegraphics[scale = 0.2, angle = 0, trim = 0 0 0 0]{Slide1.jpg}\\
%    \caption{\label{color-period}Period-color distribution of AB Dor members from \citet{Messina2011}. The target 2MASS J15594729+4403595 is marked with an red open circle.}
%\end{figure}

\indent
\indent
In our maps based on the TESS time series, we found clear evidence of a periodic variation in the spotted area, on opposite hemispheres and in the anti-phase. 
The timescale of the spot area variation is very short, with a period of about six stellar rotations, and its pattern has remained very stable during the three TESS sectors, spanning a time interval of about two years.

On the Sun, the lifetime of spots depends on the spot area according to the Gnevyshev–Waldmeier (GW) relation (\citealt{Gnevyshev1938}). However, studies extended to solar-type stars found shorter timescales  for the spot evolution compared to the solar case. From the Kepler data time series analysis, the lifetime of starspots is found in a range from 10 days to one year (e.g., \citealt{Giles2017}). Despite the difference in timescale, the physical mechanism, which after the magnetic field emersion into the photosphere leads to its diffusion, is probably the same.

{ The modulation shown by the spots on 2MASS J15594729+4403595, which is  periodic (P = 2.25\,d corresponding to six rotations), is too short to be interpreted as an active region growth and decay. In fact, this short timescale is at odds with the year-long timescale expected for the magnetic field diffusion. Rather, the change in the total spotted area observed in subsequent TESS sectors may more likely come from a modulation induced by the short-period Rossby wave, although the detailed mechanism remains elusive. }
In other words, the periodic oscillation that we observe in the maps in Figs.\,\ref{maps} and~\ref{spot_area} may be a manifestation of a Rossby wave, more commonly referred to as r-mode or Rieger-like cycles. { We notice that Rossby waves propagate in a retrograde sense in the reference frame rotating with the star, while our spot pattern is fixed in the  reference frame rotating with the $P_{1}=P_{\rm rot}=0.37$\,d rotation period. Nevertheless, we have assumed that the wave is propagating  in the deep stellar interior; thus, it is conceivable that we can see its effects on the total spotted area without any clearly induced systematic migration of the spots in longitude at the stellar surface.  Future studies could account for such a behaviour by clarifying the mechanism through which the wave affects the spot formation and location on the stellar surface. For example, the spot locations could be associated with convectively induced active longitudes that are fixed in the stellar reference frame \citep[cf.][]{Weberetal13}, while the passage of a retrograde propagating Rossby wave only induces a modulation of their levels of activity. }

To put our suggestion in context, Rieger cycles have been observed in the Sun, and only recently in solar-type stars; notably, in  CoRoT-2, which is a G7 dwarf with a period of P $\sim$ 29\,d (which corresponds to about six stellar rotations), by \citet{Lanza2009}, or on Kepler-17 by \citet{Bonomo2012}, where the cycle lasts about three stellar rotations (P$_{rot}$ = 12\,d).

What makes 2MASS J15594729+4403595 interesting is the extremely short timescale of the candidate Rieger cycle, which is more likely related to Rossby waves than to active region growth and decay or to cyclical activity. { An important characteristic of Rieger cycles is that they are not persistent, but appear close to the maxima of the solar 11-yr cycle. Also, Rossby waves have a limited lifetime in the Sun that does not exceed $\sim 1.4$ years \citep[][Sect.~4.2.2]{Zaqarashvilietal21}. Similarly, a limited lifetime for the Rossby waves in 2MASS J15594729+4403595 could account for the absence of the $P_{2}$ periodicity  in the Zeta UMa Observatory time series.}\\

\rm 
As is shown in Fig.\,\ref{TESS_figs}, the photometric modulation pattern shown by 2MASS J15594729+4403595 is easily recognisable. This offers a criterion to select other stars with similar patterns to be examined in search for Rieger-like cycles, and thus improves the total number and the statistics of this phenomenon on stars other than the Sun.

\section{Conclusion}
Our analysis has allowed us to better characterise 2MASS J15594729+4403595, which is a candidate member of the AB Dor association. Our photometric and spectroscopic observations reveal that this is  a triple system consisting of a primary M2 component, which is itself a SB1 close binary, and a secondary M8 component. We have derived  tentative  orbital parameters that may indicate \rm a system in an eccentric orbit with an orbital period of about 14\,d. We have measured the rotation period, P = 0.3701\,d, of the primary M2 component and inferred an inclination, $i$ $\simeq$ 50$^\circ$, of the stellar spin axis to the line of sight. Another periodicity of 0.44\,d could be associated with SDR or, more likely, with a Rossby wave. The spot modelling has allowed us to discover a spotted area modulation that resembles a possible Rieger-like cycle, but on a very short timescale that can be accounted for by the same Rossby wave assumed to produce the 0.44\,d periodicity in the light modulation. The extremely short candidate Rieger cycle and the multiple photometric periodicities observed in 2MASS J15594729+4403595 all make this star a very interesting target for additional studies.
 As a future perspective, we intend to extend our analysis to other stars showing similar photometric modulation patterns to explore the presence of similar Rieger cycles.

\begin{acknowledgements}
SM thanks Roi Alonso and Antonio Luis Cabrera Lavers for their contribution on the observational side. This research was funded by the European Union – NextGenerationEU" RRF M4C2 1.1 n: 2022HY2NSX. "CHRONOS: adjusting the clock(s) to unveil the CHRONO-chemo-dynamical Structure of the Galaxy” (PI: S. Cassisi) and supported by the program "Stellar activity and dynamo theory in the era of precision stellar astrophysics (P.I. A. Bonanno) within Ricerca Fontamentale at INAF. Based on observations made with the Nordic Optical Telescope, owned in collaboration by the University of Turku and Aarhus University, and operated jointly by Aarhus University, the University of Turku and the University of Oslo, representing Denmark, Finland and Norway, the University of Iceland and Stockholm University at the Observatorio del Roque de los Muchachos, La Palma, Spain, of the Instituto de Astrofisica de Canarias. 
The TESS data presented in this paper were obtained from the Mikulski Archive for Space Telescopes (MAST). STScI is operated by the Association of Universities for Research in Astronomy, Inc., under NASA contract NAS5-26555. Support for MAST for non-HST data is provided by the NASA Office of Space Science via grant NNX13AC07G and by other grants and contracts. This research has made use of the SIMBAD database, operated at CDS, Strasbourg, France. 
\end{acknowledgements}
\bibliographystyle{aa.bst} % style aa.bst
\bibliography{biblio.bib} % your references Yourfile.bib

\appendix

\section{Zeta UMa Observatory photometry}
\label{app:ZetaUMa}

To determine the stellar rotation period, we carried out a multiband photometric monitoring in 2013 at the Zeta UMa Observatory
(709 m a.s.l.; Madrid, Spain).
\begin{table*}[!h]
\caption{List of targets for differential photometry.}
\label{table:comp}
\centering
\begin{tabular}{clcccc}
\hline
        & Name & RA & DEC & Vmag & $\sigma_{\star-c}$\\
        &            & (J2000.)  & (J2000.) & (mag) & (mag)\\
        \hline
 V &      2MASS J15594729+4403595 & 15:59:47.29 &  +44:03:59.5 & 11.86 & 0.008\\
C     & TYC 3067-1461-1 	&16:00:01.86 & +44:06:01.8	& 11.37 & ... \\
CK1 & TYC 3060-1156-1	& 15 59 42.27 & +43:59:17.2 & 11.00 & 0.012\\
%CK2 & 2MASS J15593714+4355473	&15:59:37.31 &+43:55:47.6 &12.75 & 0.046\\
CK2 &2MASS J15595409+4358586 	&15:59:54.10 & +43:58:58.8&12.57 & 0.015\\
\hline
\end{tabular}
\tablefoot{Target (V), comparison (C), and check stars (CK) used for the differential photometry.}
\end{table*}

\begin{table}[!h]
\caption{Log of observations at the Zeta UMa Observatory}
\label{table:log}
\begin{tabular}{cccc}
\hline
Date & HJD$_{mean}$	& \# Frame	& \# Frame\\
        &				& V-band           & R-band\\
\hline
2013-05-31 &	2456444.49770& 149 & 12\\
2013-06-01 &	2456445.52872& 274 & 12\\
2013-06-14 &	2456458.49934 & 219	 & 12\\
2013-06-15 &	2456459.51744 & 270 & 12\\
2013-06-22 &	2456466.46756 & 45 & 12\\
2013-06-28 &	2456472.53530 & 253 & 12\\
2013-06-29 &	2456473.49245 & 177 & 12\\
2013-07-26 &	2456500.42421 & 140 & 12\\
2013-07-30 &	2456504.45789 & 203 & 12\\
2013-08-02 &	2456507.47174 & 223 & 12\\
2013-08-03 &	2456508.42086 &  166 & ...\\
2013-08-08 &	2456513.44935 & 236 & ...\\
\hline
\end{tabular}\\
\end{table}
The observations were collected by a 130mm Takahashi refractor equipped with a cooled QHY9 camera and a set of  V and R Johnson-Cousins filters.
The telescope field of view (FoV) of about 80$^{\prime}$$\times$\,60$^{\prime}$ was centred on our target star.
The observations were carried out for a total of 12 nights from May 31 to August 08, 2013. We collected a total of 2355 frames in the V and 120 in the R filter (see Table\,\ref{table:log}). The integration of 100\,s on the first night was reduced to 50\,s on subsequent nights owing to saturation of the brightest stars in the FoV. On each clear night, our target was observed continuously for about 5-6 hr together with a series of bias and flat-field frames.
%\begin{figure}
%\includegraphics[width=50mm,height=80mm,angle=90]{accuracy_V_50s.pdf}
%\caption{Photometric precision versus V magnitude with 50 sec integration measured by the task $phot$ within DAOPHOT.}
%\label{accuracy}
%\end{figure}

The data reduction was carried out using the DAOPHOT tasks within IRAF\footnote{IRAF is distributed by the National Optical Astronomy Observatory, which 
is operated by the Association of the Universities for Research in Astronomy, inc. (AURA) under 
cooperative agreement with the National Science Foundation.}. After bias subtraction and flat-fielding, we extracted the magnitudes of all stars detected in each frame using a set of different apertures. We selected the aperture giving the best photometric accuracy of our target and comparison stars. 
After removing outliers by applying a 3$\sigma$ threshold, we were left with 2251 V-band and 112 R-band measurements useful for the subsequent analysis.
We identified three stars close to the FoV centre and non-variable during the whole period of our observations which served as comparison (C) and check (CK) stars to get differential magnitudes of our target (see Table\,\ref{table:comp}). The nominal accuracy of our observations is better than 0.005 mag in the magnitude range of our target and comparison stars. However, to measure the effective photometric accuracy of our observations, instead of using the values provided by DAOPHOT and based on photon statistics, we sectioned our time series into bins of 15 minutes width (corresponding on average to 8-9 consecutive measurements) computing means and standard deviations. We found that the average standard deviation for the binned V$-$C measurements was $\sigma_{V-C}$ = 0.006 mag, whereas $\sigma_{CK1-C}\simeq\sigma_{CK2-C}$  = 0.010 mag. These values represent a more effective estimate of the  precision of our photometry (see Table\,\ref{table:comp} for the precision in the whole time series).  

\section{Spot Model}
\label{spot_model}
\subsection{Code description}
We apply a spot modelling approach already introduced in Sect. 3 of \citet{Bonomo2012}
to which we refer the reader for details. Briefly, the surface
of the star is subdivided into 200 surface elements that contain unperturbed photosphere and dark spots. The specific intensity of the unperturbed photosphere in the
TESS passband is assumed to vary according to a quadratic
limb-darkening law:
\begin{equation}
I(\mu) = I_0 (a_p + b_p\mu + c_p\mu^2)  
\end{equation}

where I$_0$ is the specific intensity at the centre of the disc, $\mu$ =
cos\,$\theta$ with $\theta$ being the angle between the local surface normal
and the line of sight, and a$_p$, b$_p$, and c$_p$ are the limb-darkening
coefficients in the TESS passband.
The dark spots are assumed to have a fixed contrast c$_s$ =
I$_{spot}$($\mu$)/I($\mu$) in the TESS passband, where I$_{spot}$ is the specific
intensity in the spotted photosphere. The fraction of a surface
element covered by dark spots is given by its filling factor $f$.
 
This model is fitted to a segment of the light curve of duration
 $\Delta$t$_f$, which is set equal to one stellar rotation period, by varying the filling factors of the individual
surface elements that can be represented as a 200-element vector
 $\vec f$. Therefore, the model has 200 free parameters and suffers from
non-uniqueness and instability due to the effect of photometric
noise. To select a unique and stable solution, we apply a maximum entropy (ME) regularisation by minimising a functional $Z$
that is a linear combination of the $\chi^2$
and of a suitable entropy
function $S$ :
\begin{equation}
Z = \chi^2({\vec f}) - \lambda S({\vec f}),    
\end{equation}
where $\lambda > 0$ is a Lagrangian multiplier that controls the
relative weights given to the $\chi^2$ minimisation and to the configuration entropy of the surface map $S$  in the solution. The expression of $S$ is given in Eq. (5) of \citet{Bonomo2012}; it is
maximal when the star is unspotted, that is, when all the elements of the vector $\vec f$ are zero. In other words, the ME criterion
selects the solution with the minimum spotted area compatible
with a given $\chi^2$ value of the best fit to the light curve. When
the Lagrangian multiplier $\lambda = 0$, we obtain the solution corresponding to the minimum $\chi^2$
that is unstable. By increasing $\lambda$,
we obtain a unique and stable solution at the price of increasing
the value of the $\chi^2$. An additional effect is that of making the
residuals between the model and the light curve biased towards
negative values because we reduce the spot filling factors by
introducing the entropy term (see \citealt{Lanza1998}; \citealt{Lanza2016}
for details).
The information on the latitude of the spots is lacking in
our maximum-entropy maps although the inclination of the stellar spin axis $i$ = 50 deg is far from being equator-on. Therefore, we limit ourselves to mapping the distribution of the filling
factor versus the longitude.

The optimal value of the Lagrangian multiplier $\lambda$ is obtained
by imposing that the mean $|\mu_{reg}| $ of the residuals between the
regularised model and the light curve verifies the relationship
(\citealt{Bonomo2012}; \citealt{Lanza2016}):
\begin{equation}
    |\mu_{reg}|  =
\sigma_0/\sqrt{N}
\end{equation}
where $\sigma_0$ is the standard deviation of the residuals of the unregularised model, that is, that computed with $\lambda = 0$, with $N$ being
the number of datapoints in the fitted light curve interval of
duration $\Delta$t$_f$.

\subsection{Testing spot modelling  results}

\begin{figure}
    \includegraphics[scale = 0.35, angle = 0, trim = 0 150 100 100]{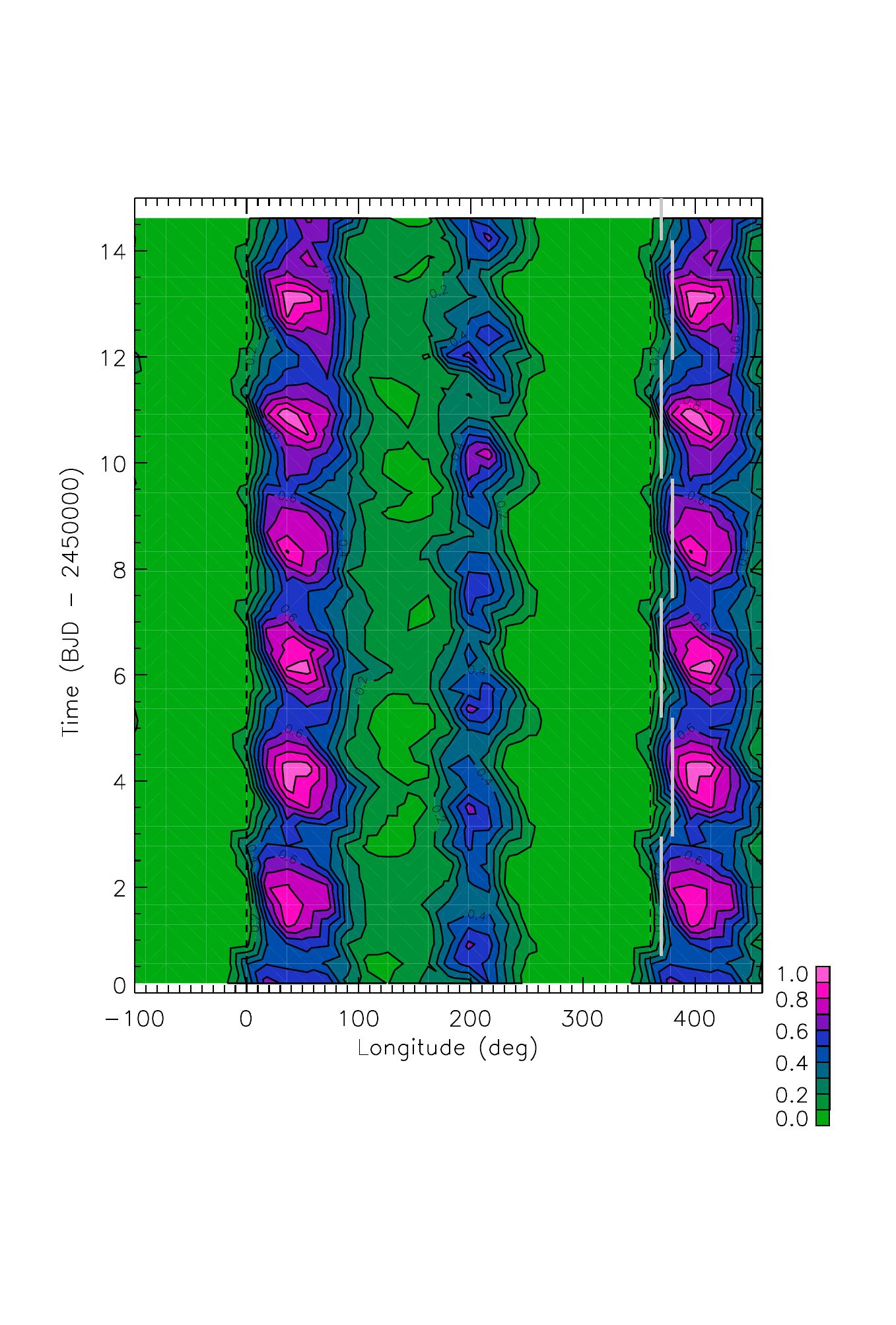}
    \caption{\label{spot_contours_argd} Map time series retrieved by modelling the synthetic light curve in Fig.\,\ref{synthetic_lightcurve_ARGD}.}
\end{figure}

{The spot pattern presented in Sect.\,\ref{sec_spot_model} may be in principle an artifact of the code, arising from its attempt to model the beat-like pattern of the observed light curve. 
To  explore \rm this possibility, as well as to discriminate between the SDR and Rossby wave interpretations presented in Sect.\,\ref{Discussion}, we performed two tests by using synthetic light curves. \\

{
Test 1 - Surface differential rotation. \rm }\\

We have combined two different synthetic sinusoids which represent the flux rotational modulations with the two periods P$_1$ = 0.37\,d (the rotation period of the major spot group) and P$_2$ = 0.443\,d (the rotation period of the secondary spot group at a different latitude). The amplitudes of the photometric modulations are similar to those observed and  a level of Gaussian noise comparable to the observed one was added:  
\begin{equation}
    flux = A_1 + B_1 \cos(\omega_1 t) +
    A_2 + B_2 \cos(\omega_2 t) + \epsilon, 
\end{equation}
where $\omega_1 = 2\pi/P_1$ and $\omega_2 = 2\pi/P_2$, $t$ is the time, and {$\epsilon$ a white Gaussian noise.}
\begin{figure}
    \includegraphics[scale = 0.35, angle = 90, trim = 0 100 0 100]{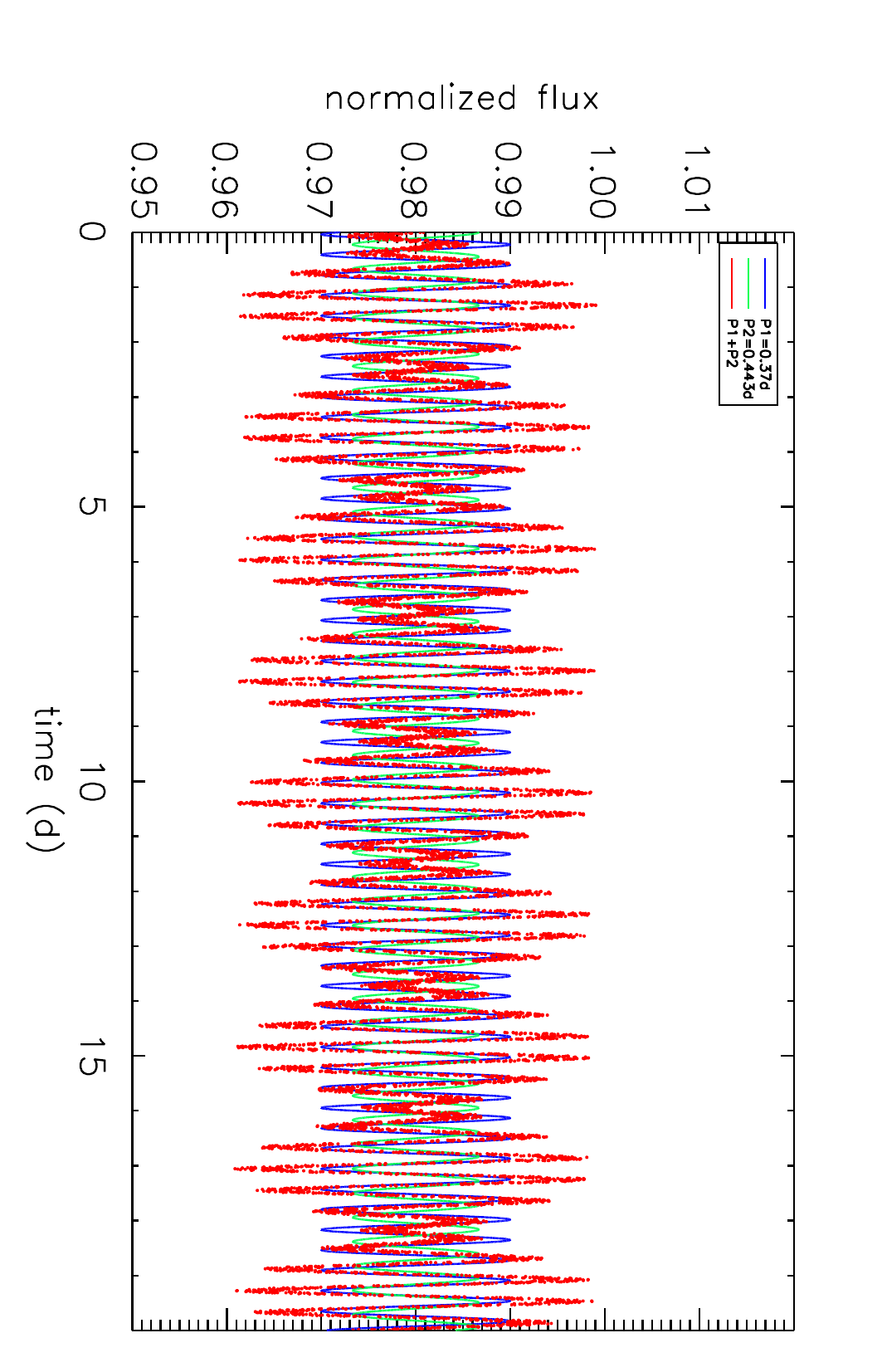}
    \caption{\label{synthetic_lightcurve} Synthetic light curve (red) obtained combining two sinusoids with $P_1 = 0.37\,$d (blue line) and $P_2 = 0.443\,$d (green line) in the Test 1 (Surface differential rotation).}
\end{figure}

We got the synthetic light curve shown in Fig.\,\ref{synthetic_lightcurve}. Then, we run our spot model on this light curve and obtained the spot map shown in Fig.\,\ref{spot_contours_synthetic}.     \\
\begin{figure}
    \includegraphics[scale = 0.35, angle = 0, trim = 0 150 100 100]{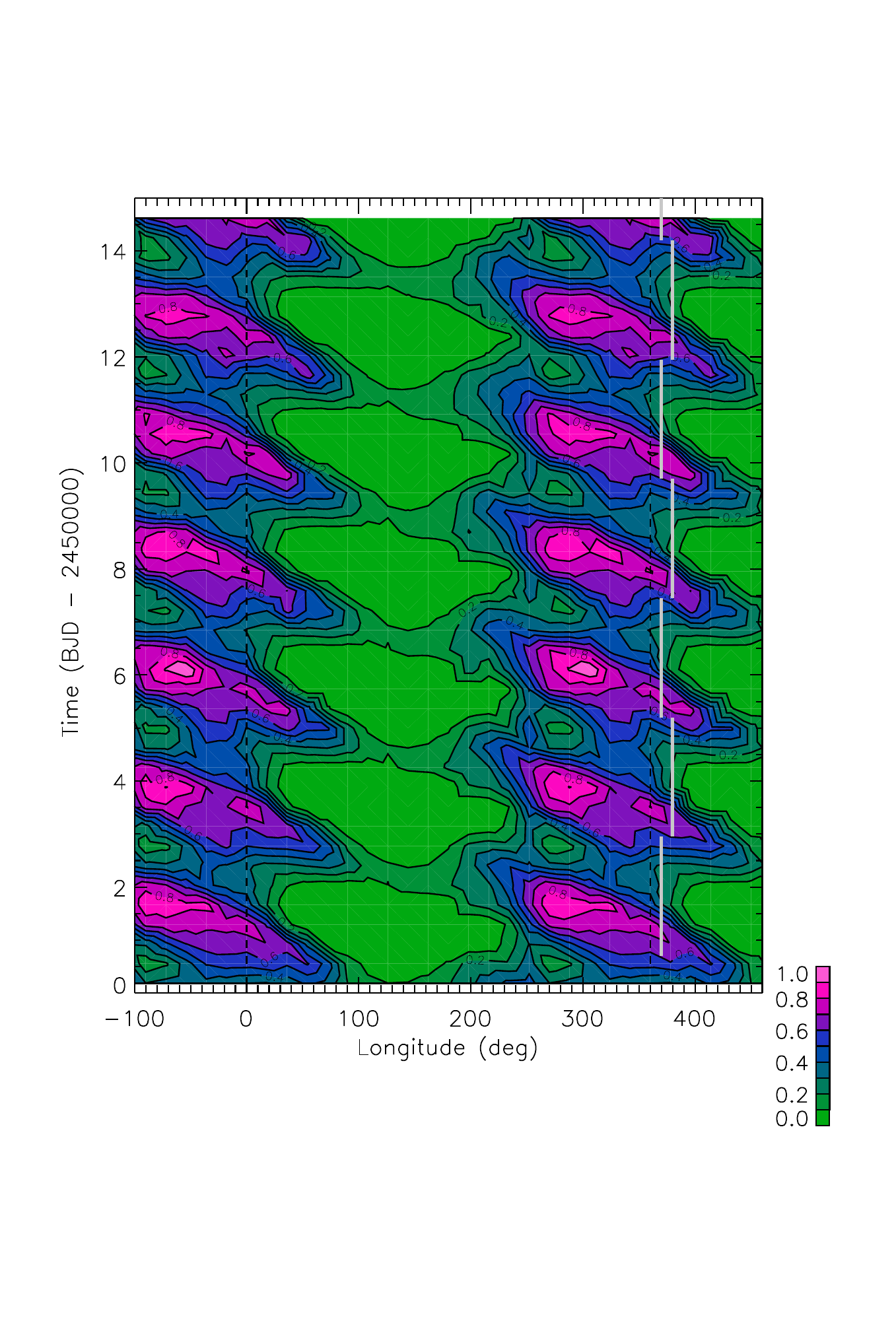}
    \caption{\label{spot_contours_synthetic} Maps time series retrieved by modelling the synthetic light curve in Fig.\,\ref{synthetic_lightcurve}}
\end{figure}
We find the activity mostly located on one longitude only with a clear drift toward decreasing longitudes with a periodicity equal to the 2.25-d beating period and with the total spotted area to remain constant from cycle to cycle.\\

{ Test 2 - Rieger cycle. \rm}\\
We have generated a synthetic light curve  which is a sinusoid with period P$_1$ = 0.37\,d (the stellar rotation period) whose amplitude is modulated by another sinusoid with a period P$_2$ = 2.25\,d, mimicking some sort of a cycle in the spottedness level. We also add the rotation period first harmonic P$_3$ = 0.1849\,d and the P$_4$ = 0.443\,d periodicity.
A Gaussian noise comparable to the observed one was added to the synthetic data and the sinusoid amplitudes were chosen to reproduce the observed ones. {Specifically, we simulate the effect of the spot area modulation with the period $P_{2}$ as}
\begin{equation}
    flux = A(t) \sin(\omega_1 t)
 \end{equation}
 where
 \begin{equation}
    A(t) = A_0 + B_0 \cos(\omega_2 t).
\end{equation}
To such a flux modulation we add the effects of the rotation and of the Rossby wave as seen by a distant observer with periods $P_{1}$ and $P_{4}$, respectively. Using the Werner formula, the  total flux \rm $f$ can be written as}
\begin{eqnarray}
f & = &A_0 \sin(\omega_1 t) +\frac{1}{2}B_0\{\sin[(\omega_1+\omega_2)t] + \sin[(\omega_1- \omega_2)t]\} \nonumber \\
& & + A_3 \sin(\omega_3 t) + A_4 \sin(\omega_4 t) + \epsilon
 \end{eqnarray} 
where $\omega_1 = 2\pi/P_1$, $\omega_2 = 2\pi/P_2$, $\omega_3 = 2\pi/P_3$, $\omega_4 = 2\pi/P_4$, $t$ is the time, and $\epsilon$ a white Gaussian noise.
\begin{figure}
    \includegraphics[scale = 0.35, angle = 90, trim = 0 100 0 100]{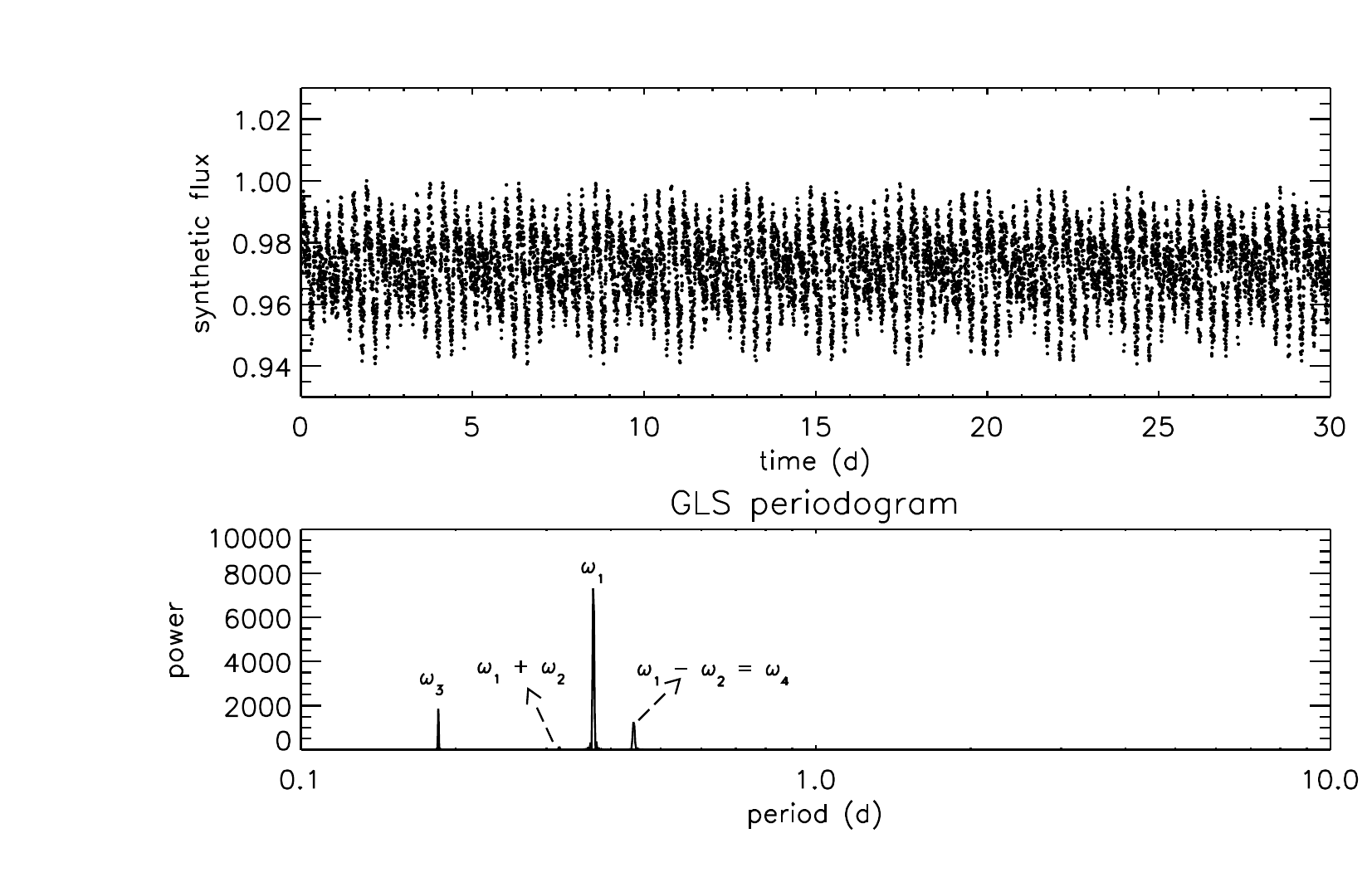}
    \caption{Periodogram analysis for Test 2 light curve. Top panel: Synthetic light curve for the Test 2. Bottom panels: the GLS periodogram where the main frequencies as detected in the observed time series are all retrieved. {The level corresponding to the white Gaussian noise added to take into account the photon shot noise is very low and does not appear in the periodogram.}}
\end{figure}
We get the synthetic light curve shown in Fig.\,\ref{synthetic_lightcurve_ARGD}. 
We modelled this light curve and obtained the  spot pattern shown in Fig.\,\ref{spot_contours_argd}.\\

If we compare the results of our Test 1 and 2 with the maps obtained from the observed data, Test 2 seems to better explain the {spot map based on the }observed data. We have two major active longitudes where the area evolves in time with a period P=2.25\,d and which is interpreted as a Rieger cycle.

A GLS {periodogram} of {Test 2 time series} recovers the main frequency $\omega_1$ as well as the additional frequencies ($\omega_1+\omega_2$) and ($\omega_1 - \omega_2 = \omega_{4}$) thus accounting for the observed periodicity $P_{4} = 0.443$\,d and indicating the Rieger cycle hypothesis as a plausible one.
We note that the frequency $\omega_1+\omega_2$ is hidden in the power background noise and becomes clearly visible only when the added Gaussian noise is removed. {Therefore, we can account for its lack in the periodogram of the TESS time series. }

\section{Effects of rapid rotation and magnetic fields on Rossby waves}
\label{app_rossby}

{ The perturbation of the frequencies of  Rossby waves given by Eq.~\eqref{rossby_freq} under the effect of stellar rotation was analysed by \citet{Provostetal81} and can be written as
\begin{equation}
\sigma \simeq \sigma_{0} \left[ 1 + \left( \frac{\Omega}{\Omega_{g}} \right)^{2} \right], 
\end{equation}
where $\sigma$ is the perturbed frequency, $\sigma_{0}$ the unperturbed frequency as given by Eq.~\eqref{rossby_freq}, $\Omega$ the rotation frequency of the star, $\Omega_{\rm g} = \sqrt{GM/R^{3}}$, $\sigma_{1}$ a coefficient depending on the wavenumbers $m$ and $n$ of the mode and the stratification of the density inside the star, $G$ the gravitation constant, $M$ the mass, and $R$ the radius of the star \citep[cf. Eq.~59 of][]{Zaqarashvilietal21}. 

We adopt a mass of our star of $M=0.4$~M$_{\odot}$ and a solar chemical composition (Z=0.02). A non-rotating stellar interior model has been computed using the Modules for Experiments in Stellar Astrophysics \citep[MESA, see][and references therein]{Paxtonetal19} assuming a ratio of the mixing length to the pressure scale height $\alpha_{\rm mlt} = 2$ and their standard network of nuclear reactions. The radius of the star once it settles on the MS is found to be  $R=0.35$~R$_{\odot}$, while the radius at the base of its convection zone is $r_{\rm b} = 0.18$~R$_{\odot}$. 

Considering a mass $M=0.4$~M$_{\odot}$,  $R=0.35$~R$_{\odot}$, and a convective stratification in the layer where the wave is propagating, we find a perturbation of the frequency of the wave of $\sim 23$\% with respect to the value given by Eq.~\eqref{rossby_freq} for the considered mode with $n=3, m=1$. This is too a large value to be compatible with the observational error. On the other hand, assuming that the wave is propagating at the base of the convection zone at $r=r_{\rm b}$, we get a perturbation of  3\%, while, by assuming propagation on top of the radiative zone, we find a perturbation of only 0.1\%. Therefore, the analysis of the rotational perturbation suggests that the Rossby wave assumed to be responsible for the $P_{2}$ periodicity is likely propagating close to the base of the convection zone or to the top of the radiative interior. 

The presence of a strong magnetic field modifies the Rossby waves by splitting the unperturbed waves into two modes with higher and lower  frequencies, respectively \citep[cf. Sect.~3.3 of][]{Zaqarashvilietal21}. We focus on the so-called fast magnetic Rossby waves as given by Eq. (76) of \citet{Zaqarashvilietal21} because the frequency of the slow mode is much smaller than the rotation frequency in our case \citep[cf. Eq.~77 of][]{Zaqarashvilietal21}. Adopting the density given by our MESA interior model for the base of the convection zone ($\rho = 30.9$~g~cm$^{-3}$) and a toroidal magnetic field of $10^{5}$~Gauss, we find a relative perturbation of the frequency of the fast mode of 0.0021\%, that is, completely negligible, because of the large predominance of the Coriolis force in our rapidly rotating star over the Lorentz force. For completeness, we estimate the frequency of the slow mode that is found to be $2\times 10^{-5}$ of the rotation frequency, that is, completely away from the range of the observed periodicities in our star. The perturbations of the Rossby wave frequencies scale with the square of the magnetic field intensity, therefore, the above conclusions are true also for magnetic fields stronger by more than one order of magnitude than the assumed field. 
}

\end{document}